\documentclass[pdflatex,sn-mathphys-num]{sn-jnl}
\usepackage{graphicx}%
\usepackage{multirow}%
\usepackage{amsmath,amssymb,amsfonts}%
\usepackage{amsthm}%
\usepackage{mathrsfs}%
\usepackage[title]{appendix}%
\usepackage{xcolor}%
\usepackage{textcomp}%
\usepackage{manyfoot}%
\usepackage{booktabs}%
\usepackage{algorithm}%
\usepackage{algorithmicx}%
\usepackage{algpseudocode}%
\usepackage{listings}%
\usepackage{url}
\usepackage{pdflscape}    
\usepackage{adjustbox}    
\usepackage{booktabs}     
\usepackage{array}    
\usepackage{subfigure}
\usepackage{natbib}
\usepackage{afterpage}
\usepackage{float}
\DeclareMathAlphabet{\mathbbold}{U}{bbold}{m}{n}

\begin{document}

\title{Redistricting from the Bottom Up: Sampling Communities of Interest with Differential Privacy}

\affil[1]{Department of Mathematics, Dartmouth College
}

\affil[2]{\orgdiv{Department of Mathematics and Statistics}, \orgname{Vassar College}
}

\author*[1]{\fnm{Atticus} \sur{McWhorter}}\email{atticus.w.mcwhorter.gr@dartmouth.edu}
\author[1]{\fnm{Caroline} \sur{Hammond}}
\author[1]{\fnm{Nianqiao Phyllis} \sur{Ju}}
\author[2]{\fnm{Daryl} \sur{DeFord}}

\date{March 2026}

\abstract{
    Independent Redistricting Commissions (IRCs) are a promising tool for bottom-up redistricting, but their public testimony processes are vulnerable to adversarial manipulation. We propose using differential privacy to draw redistricting plans that incorporate community of interest (COI) testimonies while remaining robust to adversarial input. Treating individual testimonies as data points, we use the marked edge walk (MEW) to sample from differentially private distributions of redistricting plans via the exponential mechanism. We introduce two score functions and demonstrate that both can be targeted by MEW across a range of privacy budgets. Applying this method to Missouri's mid-cycle  redistricting using 808 COI testimonies, we show that COI-informed sampling outperforms an uninformed baseline and the enacted plan. An adversarial experiment demonstrates that the method can be robust to attacks under certain privacy budgets and may perform better in practice than formal group privacy guarantees imply. We also find that stronger COI preservation tends to spread minority and Democratic representation more evenly across districts. }

\maketitle

\section{Introduction}
\label{sec:intro}
In many states, redistricting can be thought of as a top-down process; state legislatures draw the districts from which they are elected. In these cases, the process can be controlled by the majority party, rather than through a grassroots, local, or nonpartisan effort. In some cases, this approach leads  to conflicts of interest and gerrymanders, where governing bodies draw maps that unfairly advantage (or disadvantage) a particular racial or ethnic group, incumbent politician, or political party. As techniques for creating potentially unfair districting  plans have become more computational and data-driven,   developing and applying quantitative methods to determine whether a given map is gerrymandered has become an increasingly important  problem, both in academic work and in litigation.   

\medskip

Over the past decade, the ensemble method, where researchers generate large collections of feasible comparison maps, has become a prominent tool for evaluating districting plans \cite{aftermath}. Recent mathematical advances in these algorithms allow researchers to target distributions on observables of interest, and the increased availability of these methods (through packages like \texttt{gerrychain} and \texttt{redist}) has led to the adoption of these algorithms by experts in a wide range of court cases. In these cases, ensembles generated by only considering or targeting criteria derived from legislation are often used to provide evidence of intent: if a proposed map is an ensemble-outlier on one or many observables, the map-drawer may have considered external (potentially nefarious) objectives beyond just following the traditional districting criteria. In a recent example, Texas Republicans argued in court that a new congressional plan was not drawn to disenfranchise black and latino voters, rather to create five new Republican districts. In response, Professor Moon Duchin used the ReCombination algorithm to generate an ensemble of comparison plans and showed that the racial outcomes of the proposed plan were an outlier even among plans with similar Republican advantage to the enacted plan \cite{Duchin_TX}. While ensemble methods are an effective diagnostic tool for detecting top-down manipulation, they can also be applied to analyze other aspects of redistricting. As an example, Colorado's Independent Legislative Redistricting Commission hired several mathematicians as consultants to construct ensembles to help evaluate their proposals during the line drawing process \cite{COCOMM} Based on this analysis and their prior work performing ensemble analysis in the state \cite{colorado}, the commission was able to use the ensemble analysis to defend their maps before Colorado's State Supreme Court. Additional ways in which quantitative techniques could be applied to support line drawing commissions were considered in \cite{ERZ}. 

\medskip

Independent Redistricting Commissions (IRCs), such as those present in Arizona, Colorado, and Michigan, represent a more bottom-up approach to redistricting. In these states, an independent body of commissioners is chosen to insulate the legislature and political parties from the map-drawing process. To eliminate incumbency-related conflicts of interest, IRC commissioners are commonly required to not hold or run for public office for a certain amount of time. To reduce the amount of partisan influence on the process and decrease political gerrymandering, IRCs are usually constructed in a politically balanced manner with an even number of Democrats and Republicans. For example, in Colorado the commission consists of 12 individuals split between four Republicans, four Democrats, and four unaffiliated voters, and a plan must be selected by a super majority that includes at least two unaffiliated voters. Lastly, to increase transparency in the redistricting process, IRCs are often required to have a minimum number of public hearings, publish any and all meeting data, and allow ample time for responses to proposed plans. The potential and impacts of this focus on transparency has been considered by legal scholars, identifying both success and potential pitfalls \cite{RG1, RG2}. One common function of public hearings is to gather evidence about Communities of Interest (COIs), which are geographic regions that some states attempt to keep together within districts in order to allow groups with similar interests the ability to advocate for themselves to their representative. 

\medskip

Despite these precautions, bipartisan redistricting commissions are not infallible, and simply creating an IRC does not guarantee redistricting success. Examples in the current cycle include situations in Ohio, New York, and Missouri where partisanship led to deadlocks and controversy. Even when the IRC is successful in creating and passing a map there is often significant litigation surrounding the selected plan. Beyond the issues raised by lawsuits, there are also other ways that partisans may attempt to subvert the redistricting process as carried out by IRCs, including manipulating or coordinating testimony around COIs.  

\medskip

In the 2010 redistricting cycle in California, the Democratic party organized a group of voters, elected officials, and people with strong democratic party affiliation to testify before the Citizen's Redistricting Commission in support of plans that enshrined gains for the Democratic party \cite{PP_CA}. One of these testimonies came from a lobbyist who grew up in Idaho and lived in Sacramento who claimed to represent the Asian American population in the San Gabriel Valley. The report from California's 2021 commission explicitly discusses the potential for scripted COI submissions \footnote{\url{https://wedrawthelines.ca.gov/wp-content/uploads/sites/64/2023/06/mh-2023-06-26-27-RRR-Rpt-Volume3-v3.pdf}}. Also in the 2010 cycle, Republicans operatives in Florida attempted a similar strategy as described by the Supreme Court of Florida quoting the trial court \cite{FL_SC}:

\begin{quote}
``The strategy they came up with, according to the
[challengers], was to present to the public a redistricting process that
was transparent and open to the public, and free from partisan
influences, but to hide from the public another secretive process. In
this secretive process, the political consultants would make
suggestions and submit their own partisan maps to the Legislature
through that public process, but conceal their actions by using proxies,
third persons who would be viewed as “concerned citizens,” to speak
at public forums from scripts written by the consultants and to submit 
proposed maps in their names to the Legislature, which were drawn
by the consultants.

What is clear to me from the evidence, as described in more
detail below, is that this group of Republican political consultants or
operatives did in fact conspire to manipulate and influence the
redistricting process. They accomplished this by writing scripts for
and organizing groups of people to attend the public hearings to
advocate for adoption of certain components or characteristics in the
maps, and by submitting maps and partial maps through the public
process, all with the intention of obtaining enacted maps for the State
House and Senate and for Congress that would favor the Republican
Party.''
\end{quote}

\medskip

Similar attempts to cause IRCs to adjust their line drawing priorities to favor COIs have also occurred in other states. Thus, there remains a challenge for IRCs in creating a testimony process that is robust to these types of adversarial attacks. Additionally, even states that do not have IRCs often hold hearings to gather similar COI data and could be susceptible to the same types of attacks. In this paper, we propose a new point of view on bottom-up redistricting through differential privacy (DP). Differential privacy is a mathematical framework that protects individual privacy when releasing statistics by adding calibrated noise to computations, preserving patterns while limiting what can be inferred about any single data point. We provide a slightly different reading of DP here by seeking to protect the output distribution from adversarial data points, instead of protecting single data points from an overly-descriptive output. In other words, we propose a method for drawing plans that takes into account single voter testimonies without overly prioritizing any one testimony, adding robustness to the type of attacks observed in California, Florida, and elsewhere.  

\section{Background}

\subsection{Communities of Interest}

The preservation of COIs in redistricting plans is often considered one of the traditional districting criteria and approximately half of the states have a rule, either in their constitution, legislation, or guidance, that addresses COIs in redistricting \cite{BCCOI}. An overview of the complexities involved in defining this concept and the various ways that it has been operationalized is presented in \cite{Rosenfeld2022} and a geographer's account of the concept is explained in \cite{Nelson2022}. As these chapters explain, there is not significant agreement on the concept of how exactly a COI should be defined, or how data on COIs can be effectively collected and described quantitatively. The Metric Geometry and Gerrymandering Group has been particularly active in gathering data for this type of analysis, collaborating with commissions and governments in several states to use its Districtr platform to solicit COI input from a variety of stakeholders\footnote{\url{https://mggg.org/cois}}. In addition to their public-facing reports, they have also developed new methodology for  analyzing this type of data \cite{ACM}.

\medskip

Analysis of COIs in the context of redistricting goes back decades, and the ``nebulous'' nature of the concept \cite{Malone}, means that there have been many distinct proposals about how to quantify COIs for practical use. Particularly in the past decade, as computational methods have become more easily accessible, there has been a rapidly developing literature on gathering and evaluating COIs. In \cite{Chen_COI} a comprehensive analysis of previous legal and practical measures is presented, along with two new metrics called the Effective Splits Index and the Uncertainty of District
Membership, which are intended to quantify how much a COI has been split by a given plan. Other research has focused on how to identify COIs from data sources using quantitative techniques to study ballot initiatives \cite{Maske}, contact graphs \cite{Syme}, and cluster analysis \cite{Swanson}. This work has not only taken place in the academic literature, as experts have relied on these new data-focused approaches to evaluate maps in litigation \cite{DeFord_Wright, Duchin_WI}. 

\medskip

Our approach in this paper is distinct from either of these previous perspectives, in that our focus is on using techniques from differential privacy, combined with ensemble analysis, to understand the properties of maps that offer privacy guarantees and to combat potentially adversarial testimony. In addition to problem formulation and evaluation on a synthetic model, we also focus on a case study of empirical COI data gathered in Missouri at the beginning of this decade \cite{mggg_missouri_coi} and use it to evaluate the recently enacted mid-cycle plan as displayed in Figure \ref{fig:enacted} below. While there was an ensemble and optimization analysis combining ReCombination with a local search method to evaluate interactions between the state's redistricting criteria and measures of partisan fairness \cite{MO} it did not consider COI data, except for county preservation. Thus, our analysis offers new insights into the interaction between COIs and the political geography of the state. 


\subsection{Ensemble Methods}


Ensemble methods leverage the tools of Markov Chain Monte Carlo (MCMC) to sample large numbers of valid redistricting plans. In these algorithms, states are encoded as graphs where vertices correspond to the fundamental redistricting units (e.g. census blocks or voting precincts), with edges connecting adjacent units. The redistricting problem is thus reduced to graph partitioning under certain constraints, including  population balance and contiguity. See Section \ref{sec:formal} below for a formal description of this modeling process. 

\medskip

Early algorithms for this problem drew inspiration from statistical mechanics \cite{Fifield}, introducing a Glauber dynamics-inspired proposal mechanism that reassigns individual vertices to neighboring partitions \cite{Chikina_PNAS}. This local approach can exhibit slow mixing due to its small, localized steps and the presence of insurmountable energy barriers in the state space \cite{najt2019complexity,najt2020empirical}.
The search for a walk that could take larger steps motivated the development of Recombination (ReCom) \cite{DeFord2021Recombination}. Drawing inspiration from biological DNA recombination, ReCom selects two adjacent districts, merges them, and partitions the merged region into two new districts. The larger step enables the walk to traverse the energy barriers that impeded the flip walk. Combined with its implementation in the Python package \texttt{gerrychain}, ReCom has become a common algorithm of choice  for practitioners, including in expert witness testimony in court cases all over the country. However, due to the its relative complexity, the stationary distribution of ReCom is unknown. This limitation is particularly significant for applications that require an explicitly specified target distribution, for instance, when targeting policy-based distributions or implementing DP via the exponential mechanism (as we do in this paper). 

\medskip

Several related algorithms have attempted the challenge of sampling from a known stationary distribution. Reversible Recombination (RevReCom) \cite{cannon2025spanningtreesredistrictingnew} tweaks the transition probabilities of ReCom to satisfy the detailed balance for the spanning tree distribution, which weights each partition by the product of the spanning tree counts of its parts. Metropolized Forest Recombination (MFR)  operates on a lifted state space of spanning forests rather than directly on partitions, and uses a ReCom-style merge-split proposal within the Metropolis-Hastings framework \cite{MFR}. Both of these methods are very effective at sampling from distributions that place high weights on plans where the districts have many spanning trees but not all distributions of interest take this form and determining which distributions can be effectively sampled is an active area of research.  


\medskip

Two recently developed lifted walks have provided evidence of an ability to sample from distributions that distance the sampling distribution from spanning tree dependence: the cycle walk \cite{deford2025cyclewalksamplingmeasures} and the marked edge walk (MEW) \cite{mcwhorter2025marked}. Both employ similar proposal mechanisms on lifted state spaces. The cycle walk operates on the same spanning forest space as MFR, and MEW walks on tuples of a spanning tree and set of edges marked for removal. While removing the marked edges yields spanning forests, MEW maintains the marking information across transitions. Both walks take small and large steps: the cycle walk through a tuning parameter and MEW through probabilities inherent to the dual graph structure. The cycle walk demonstrates mixing from distributions with a reduced spanning tree bias compared to MFR, and MEW has shown convergence to distributions entirely independent of spanning tree structure on small graphs and to Gaussian distributions over cut edge counts on large graphs. Another recently proposed method, the Balanced Up Down Walk (BUD) \cite{akitaya2026balancedupdownwalk}, uses a similar approach to move around the space of trees and seems likely to offer more theoretical tractability than some of the previous methods.

\medskip
 
For our  application, we need to be able to specify as a target distribution the exponential mechanism. While the exponential mechanism requires only proportionality and could therefore be used with the spanning tree distribution, we adopt MEW for its demonstrated ability to sample Gaussian distributions over cut edges. 

\medskip

Alternative sampling approaches also merit consideration. Multiscale methods \cite{MS1, MS2} offer a promising new approach, addressing the mixing problems of the flip walk by merging geographic units at different levels and parallel tempering. These algorithms have demonstrated mixing when sampling from policy-based distributions and would be suitable for our application as well. We use MEW primarily for practical considerations: an existing \texttt{julia} implementation that is effective for our purposes is readily available. Our work represents an early application of MCMC-based redistricting algorithms for sampling from a specified target distribution outside of the context of evaluating specific plans.



\medskip

\subsection{Differential Privacy}
\label{sec:DP}

Differential privacy (DP, \cite{dwork2016calibrating}) is a mathematical framework for protecting individual-level information in statistical analysis and machine learning. 
Privacy of a randomized algorithm is defined in terms of the effect of the inclusion or exclusion of any single data point on the output of an algorithm. The effect is typically controlled by privacy parameters $(\varepsilon, \delta)$ with $\varepsilon > 0, 0 \leq \delta < 1.$ 

\medskip

Here, we introduce the definition of $(\varepsilon,\delta)$-DP in the context of defending against hostile testimonies. Let $\textbf{t} = \{t_i\}_{i = 1}^T$ be a set of testimonies, and let  $\textbf{t}'$  be a neighboring set, in the sense of having set difference cardinality 1, i.e. $|\textbf{t} \setminus \textbf{t'}| = 1$. A randomized algorithm $\mathcal{A}$ is said to be $(\varepsilon,\delta)$-DP, if for all such neighboring datasets, we have 
\begin{equation}
    \label{eqn:epsilon-delta-dp}
    \mathbb{P}(\mathcal{A}(\textbf{t}) \in B) \le \mathbb{P}(\mathcal{A}(\textbf{t}') \in B) \cdot e^{\varepsilon} + \delta
\end{equation}
for all measurable subsets $B$ in the output space of the algorithm, which in our context is the space of eligible redistricting plans. We delay the complete setup of the notation to Section~\ref{sec:formal}. 
Imagine $\textbf{t}, \textbf{t'}$ differing by one adversarial testimony and sampling a redistricting plan with algorithm $\mathcal{A}$. Eq.\eqref{eqn:epsilon-delta-dp} ensures that this adversarial attack only has a limited, quantifiable impact on our redistricting plan.  

\medskip

We shall use the exponential mechanism \cite{mcsherry2007mechanism} to produce a private redistricting plan. With a nonnegative score function $s(\textbf{t})$, the Exponential mechanism samples according to the (unnormalized) density/mass function
\begin{equation}
    \label{eqn:exponential-mechanism}
    q(\mathcal{A}(\textbf{t})) \propto \exp\left(- \frac{\varepsilon}{2\Delta(s)} s(\textbf{t}) \right) \cdot m(\textbf{t})
\end{equation}
where 
\begin{itemize}
    \item $\Delta(s)$ is a (global) sensitivity that describes the maximum change in the score induced by changing any $\textbf{t}$ to a neighboring $\textbf{t'}$;
    \item $m(\cdot)$ is a base measure. For example, the uniform distribution on all eligible redistricting plans would have $m(\textbf{t}) \propto 1$.
\end{itemize}

Generating $\mathcal{A}(\textbf{t})$ according to Eq.\eqref{eqn:exponential-mechanism} achieves $(\varepsilon,0)$-DP.
Since directly sampling from a distribution like Eq.\eqref{eqn:exponential-mechanism} is intractable, we design an MCMC algorithm to achieve this. 
The following has been noted in \cite{chaudhuri2013near,bertazzi2025differential}: 
if we can guarantee that the sampler’s distribution has total variation distance (TVD) $\delta$ from the exponential mechanism target,
then the sampler can guarantee $(\varepsilon,\delta (e^{\varepsilon} +1 ))$ differential privacy.


\section{Materials and Methods}

\subsection{Data}


Data for the dual graphs were acquired from the eveomett-states github repository\footnote{\url{github.com/eveomett-states}}, which are described in \cite{agarwal2025statesdisarraycleaningdata}. The dual graph is built from the 2020 census block groups, whose geometries were acquired from the Redistricting Data Hub\footnote{\url{redistrictingdatahub.org}}. 

\medskip

MGGG gathered COI testimonies in partnership with the Fair Maps Missouri and OPEN Maps Coalitions during Missouri's 2021 redistricting cycle \cite{mggg_missouri_coi}. Missourians were invited to submit their testimonies through the Fair Maps Missouri Redistricting Portal on Districtr\footnote{\url{https://portal.missouri-mapping.org/}}. The testimonies are publicly available on MGGG's github page \cite{MO_GH}, and can be downloaded from Districtr using the provided links. The portal collected 808 total submissions between June and August 2021. Each submission consists of a spatial component, a highlighted selection of census blocks defining the community boundary, and a narrative component describing the community's shared interests and why it should be kept together in redistricting. Each spatial component is comprised of the 2010 census blocks. To use these with our 2020 block group dual graph, we created a mapping using GeoPandas's spatial intersection capability.. When a 2010 block overlapped multiple 2020 block groups, we assigned it to the block group with the largest area of intersection. Each testimony was sorted into geographic clusters that can be found in the COI report \cite{mggg_missouri_coi}.
\medskip

\begin{figure}[!htbp]
    \centering
    \subfigure[]{\includegraphics[width=0.49\textwidth]{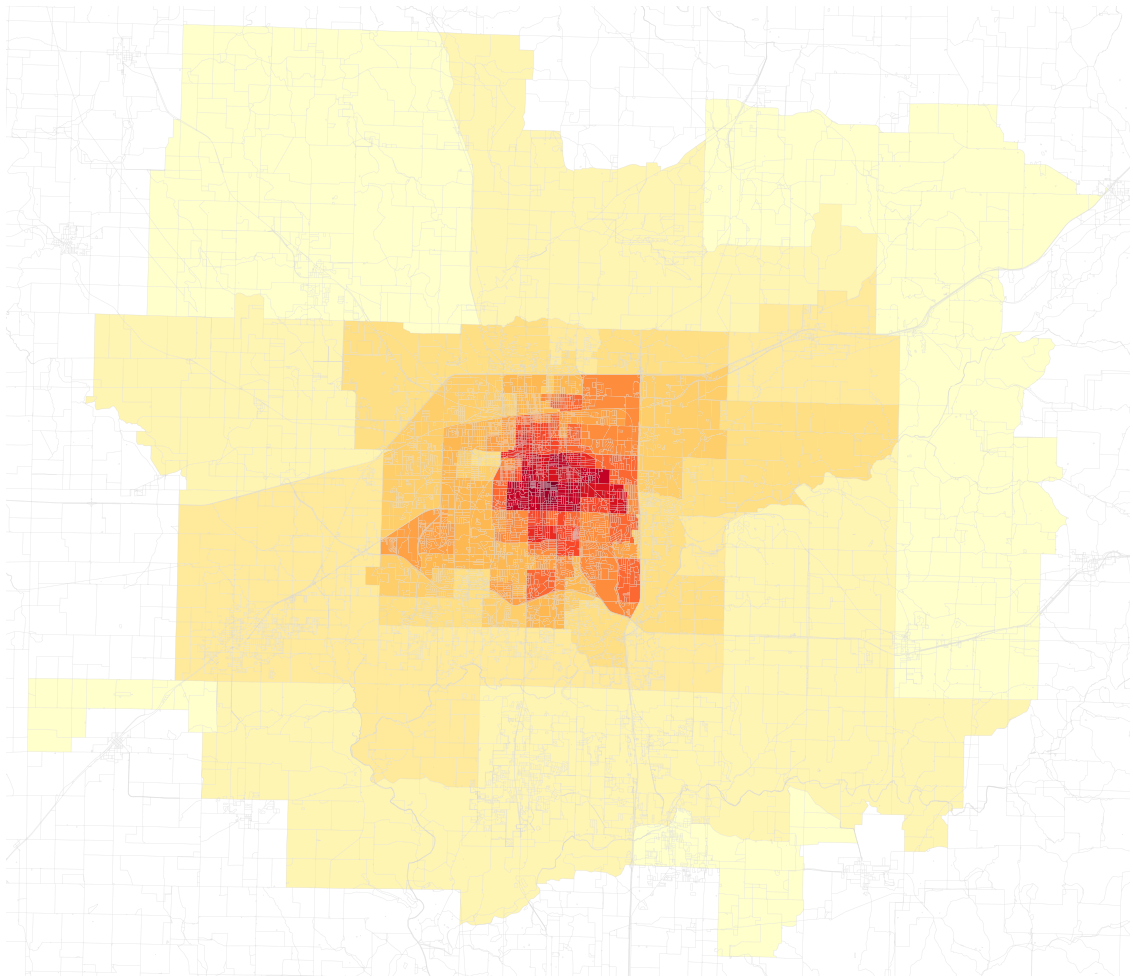}}
    \subfigure[]{\includegraphics[width=0.49\textwidth]{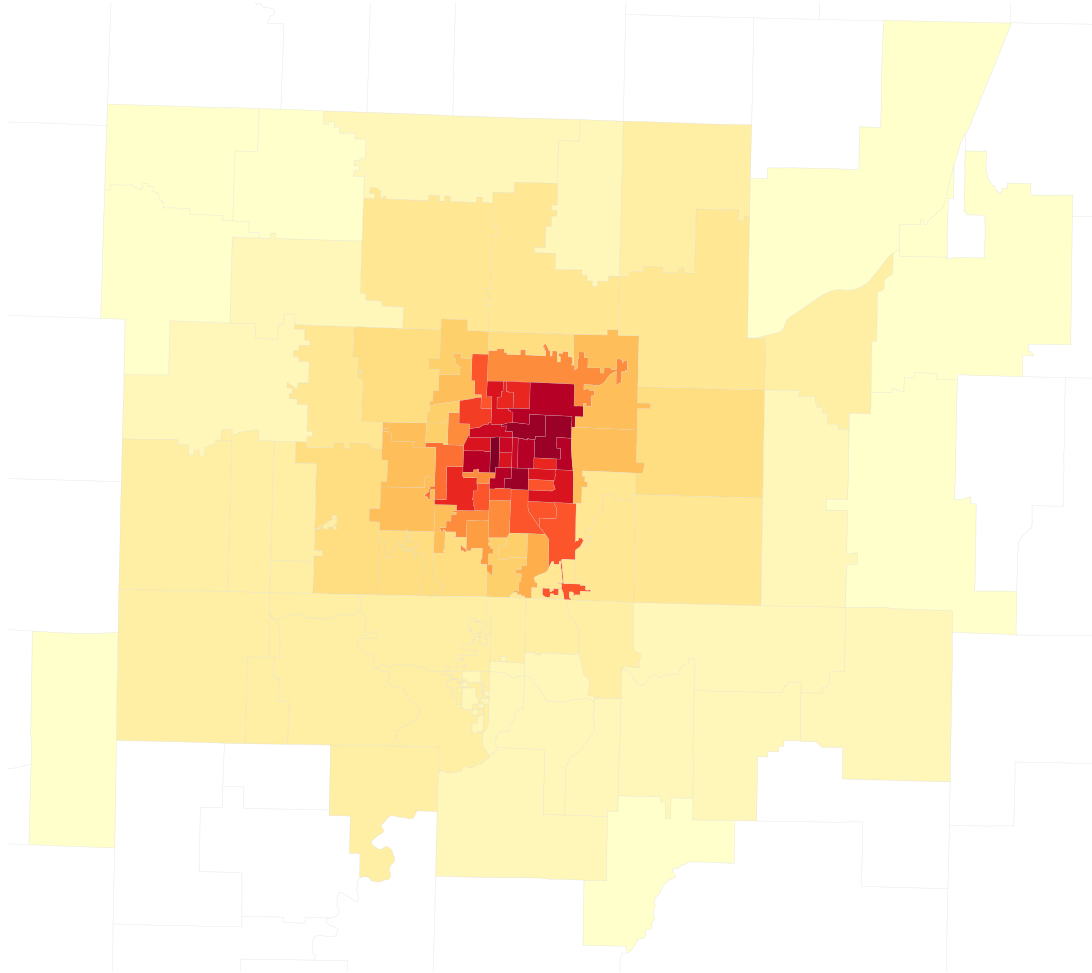}}
    \caption{\textbf{An Example COI Cluster: }Downtown Springfield's cluster shown over census blocks (a) and block groups (b), showing that aggregation to the block group level preserves the structure of the cluster.}
    \label{fig:cluster_27}
\end{figure}

Examples of a COI cluster are shown in Figure \ref{fig:cluster_27}. The cluster (C27 in \cite{mggg_missouri_coi}) represents a group of testimonies describing downtown Springfield, which contains Missouri State and two large hospitals. An example testimony in this cluster reads, ``This is the historic heart of Springfield. It encompasses 3 of our universities - Drury, OTC, and MSU, along with the downtown region with Park Central Square, a good portion of the ``Route 66", and the Springfield Art Museum. People who live, go to school, and work in this area tend to be more progressive, inclusive, and diverse". Figure \ref{fig:cluster_27} (a) displays the cluster in the form it was collected, over census blocks, while Figure \ref{fig:cluster_27} (b) displays the same cluster over block groups, demonstrating that the aggregation of blocks into block groups does not substantially change the testimonies or clusters. 

\subsection{Problem Formalization}
\label{sec:formal}




On a graph $G = (V, E)$, a redistricting plan is a partition of the vertices into $K$ connected components, and each plan is characterized by an assignment function $\xi: V \rightarrow \{1, 2, ..., K\}$. In Missouri, $K = 8$. We enforce that each part is population balanced, i.e., for each part $d$, $$\sum_{v \in \xi^{-1}(d)}\text{pop}(v) \in [\frac{P}{K} (1 - \epsilon_{\text{pop}}), \frac{P}{K} (1 +\epsilon_{\text{pop}})]$$ where $\text{pop}(v)$ is the population of node $v$, $P = \sum_{v \in V}\text{pop}(v)$ is the total population, and $\epsilon_{\text{pop}}$ is the imbalance tolerance. 

\medskip

Define the testimony space as $\mathcal{X}_t = [0, 1]^{|V|}$. Each COI testimony is encoded as a binary vector $t_i \in \mathcal{X}_t$ for $ i \in \{1, 2, ..., T\}$ where the $k$th entry is 1 if the testimony $i$ highlights block group $k$, and 0 otherwise. Here $T = 808$ is the number of testimonies. MGGG grouped each testimony into $C = 35$ clusters based on geographic proximity. Let  the cluster space be denoted by $\mathcal{X}_c = \mathbb{N}^{|V|}$. We encode each cluster as a vector $c_j \in \mathcal{X}_c$ for $ j \in \{1, 2, ..., C\}$ where the $k$th entry tallies the number of testimonies in the cluster $j$ that highlight block group $k$: $$c_j = \sum_{i \in \mathcal{I}_j}t_i$$ where $\mathcal{I}_j \subseteq \{1, ..., T\}$ is the index set of testimonies assigned to cluster $j$. 

\medskip

Given a set of testimonies $\textbf{t} = \{t_i\}_{i = 1}^T$ and a redistricting plan $\xi: V \to \{1, ..., K\}$, we define the testimony split score $s_t(\textbf{t}, \xi)$ as the proportion of testimonies that remain whole when the redistricting plan is applied to the state:
$$s_t(\textbf{t}, \xi) = \frac{1}{T}\sum_{i=1}^T \mathbbold{1}\{\xi(v) = \xi(v') \text{ for all } v, v' \in \text{supp}(t_i)\}$$
where $\text{supp}(t_i) = \{k \in V : (t_i)_k = 1\}$ is the support of testimony $i$ (the set of block groups highlighted by the testimony). The $\ell_1$-sensitivity of $s_t$ with respect to a single testimony is $\Delta(s_t) = \frac{1}{T}$. To see this, observe that for a fixed redistricting plan $\xi$, changing a single testimony $t_i$ can affect at most one term in the sum $\sum_{i=1}^T \mathbbold{1}\{\cdot\}$. Since each indicator contributes either 0 or 1, changing one testimony changes $s_t(\textbf{t}, \xi)$ by at most $\frac{1}{T}$.

\medskip

While $s_t$ provides an intuitive measure of COI preservation, it has several limitations. If a testimony highlights a region with population exceeding $\frac{P}{K}(1 + \epsilon_p)$, it must be split by any valid redistricting plan, contributing 0 to $s_t$ regardless of how it is split. Additionally, a testimony split across two districts contributes the same (0) as one split across five or six districts, despite the latter being a worse outcome. With these criteria in mind, we can begin to define another score function. Given a set of testimonies $\textbf{t}$, a set of clusters $\textbf{c} = \{c_j\}_{j = 1}^C$, and a redistricting plan $\xi$, we define the cluster weight score: 

$$s_c(\textbf{t}, \textbf{c}, \xi) = \frac{1}{W}\sum_{j=1}^C \max_{d \in \{1, ..., K\}} w_j^d(\xi)$$ where $w_j^d(\xi) = \sum_{v \in \xi^{-1}(d)} (c_j)_v$ is the weight of cluster $j$ in district $d$, and $W = \sum_{j=1}^C \|c_j\|_1$ is the total weight. In other words, for each cluster $c_j$, we identify the district $d$ that contains the largest total weight of that cluster (measured by summing $(c_j)_v$ over all block groups $v$ assigned to the district $d$). The score $s_c$ is the fraction of total cluster weight that is preserved in the `best' district for each cluster. Since the cluster vector can be obtained from the testimony vector and a (fixed) assignment function, we will often refer to the cluster score as $s_c(\textbf{t}, \xi)$. 

\medskip


The $\ell_1$-sensitivity of $s_c$ with respect to a single testimony is bounded by $\Delta(s_c) \leq \frac{|V|}{|V| +W} - \frac{1}{W}$. To see this, adding or removing a single testimony from cluster $j$ changes $(c_j)_v$ by at most 1 for each $v$. Then, the maximum weight $w_j^d(\xi)$ can change by at most $|V|$, yielding a change in $s_c$ of at most $\frac{|V|}{|V| +W} - \frac{1}{W}$. This is a conservative estimate in two ways. Firstly, testimonies tend to highlight far fewer block groups than $|V|$; on average, testimonies highlight 372 census block groups, which is roughly 10\% of the 3733 total census block groups. Secondly, due to population balance constraints, the redistricting plan will split any testimony with population greater than $\frac{P}{K}(1 - \epsilon_{\text{pop}})$.

\begin{figure}[!htbp]
    \centering

    \subfigure[]{\includegraphics[width=0.49\textwidth]{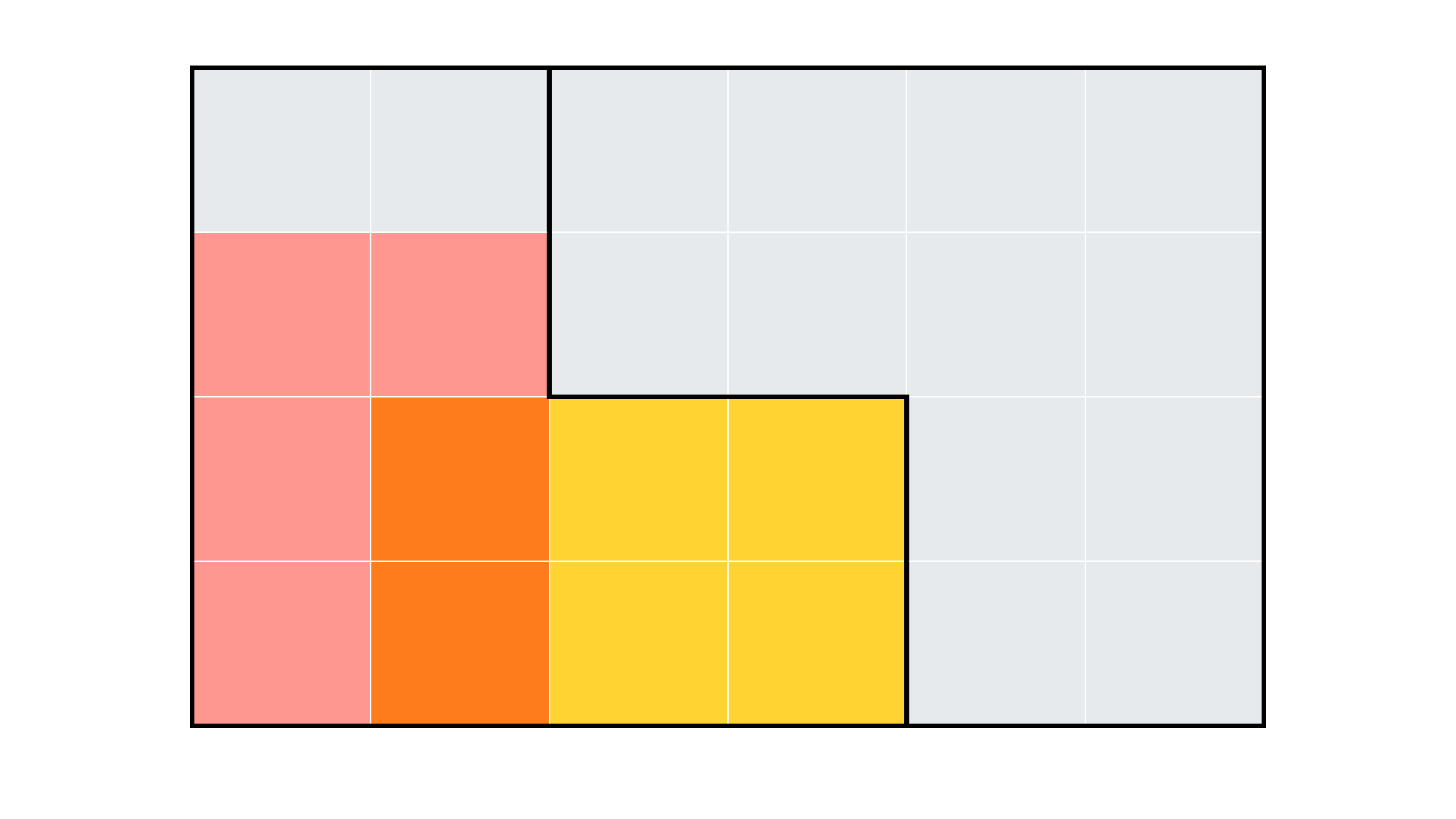}}
    \subfigure[]{\includegraphics[width=0.49\textwidth]{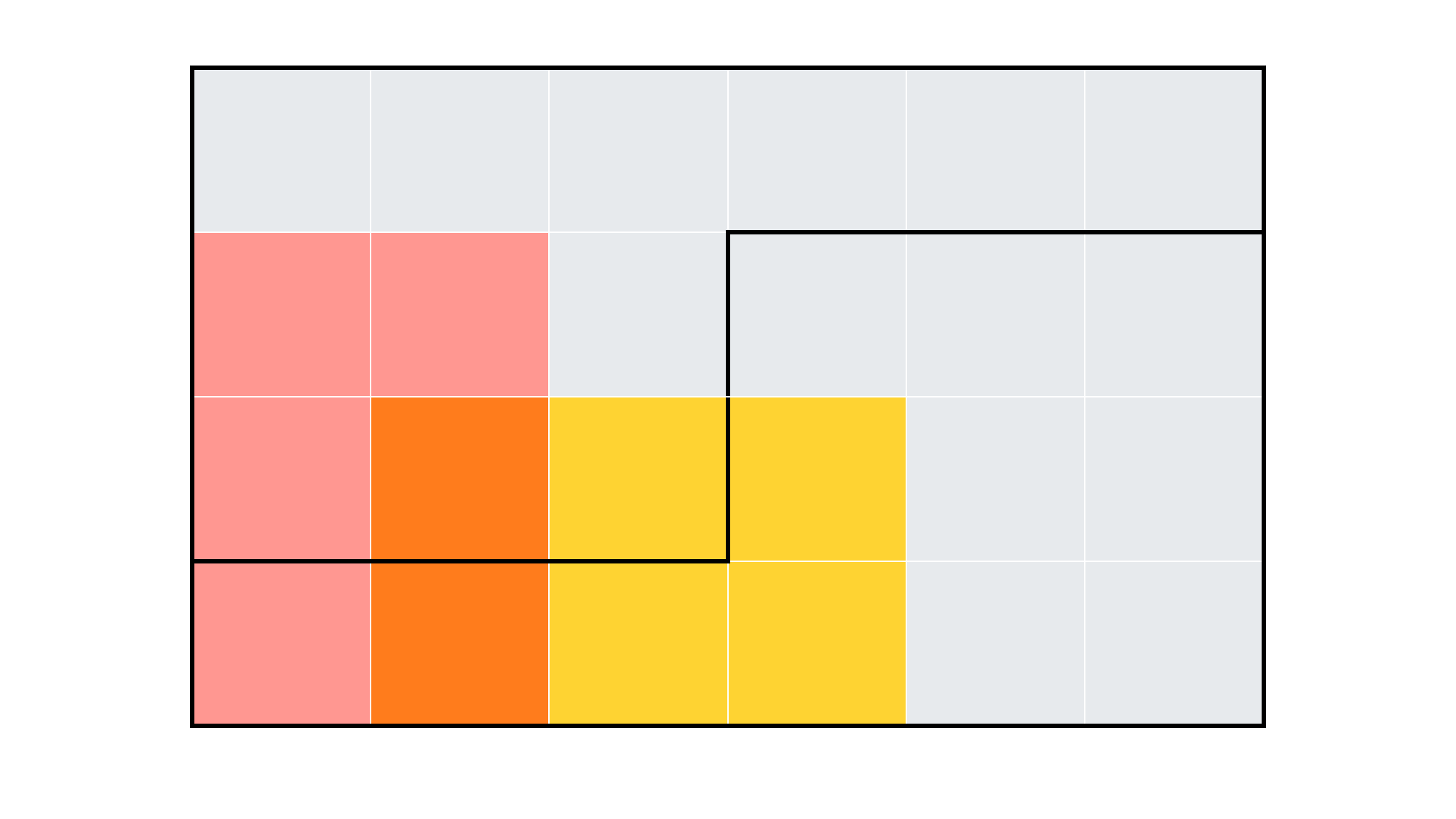}}
    \subfigure[]{\includegraphics[width=0.49\textwidth]{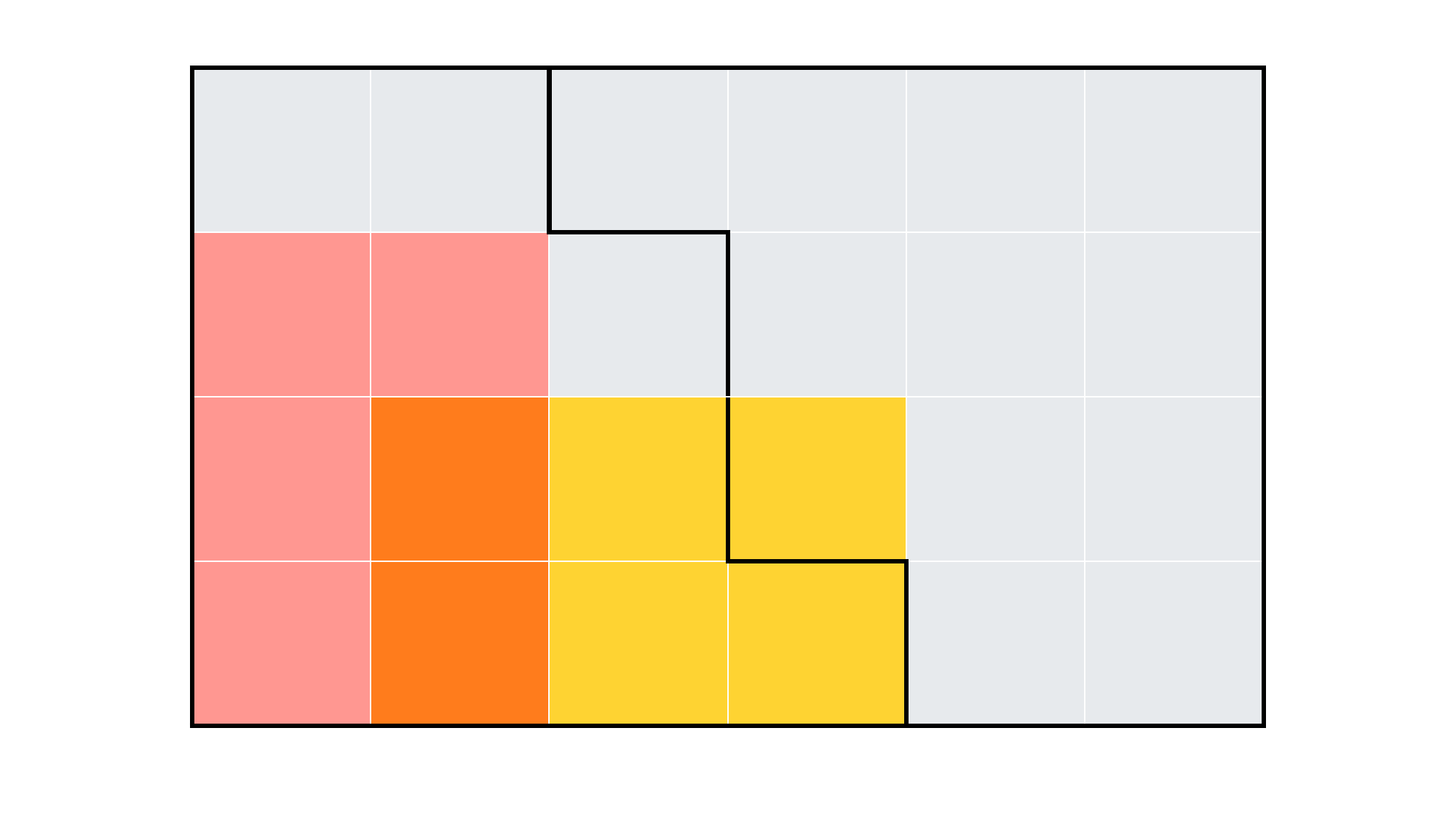}}
    \subfigure[]{\includegraphics[width=0.49\textwidth]{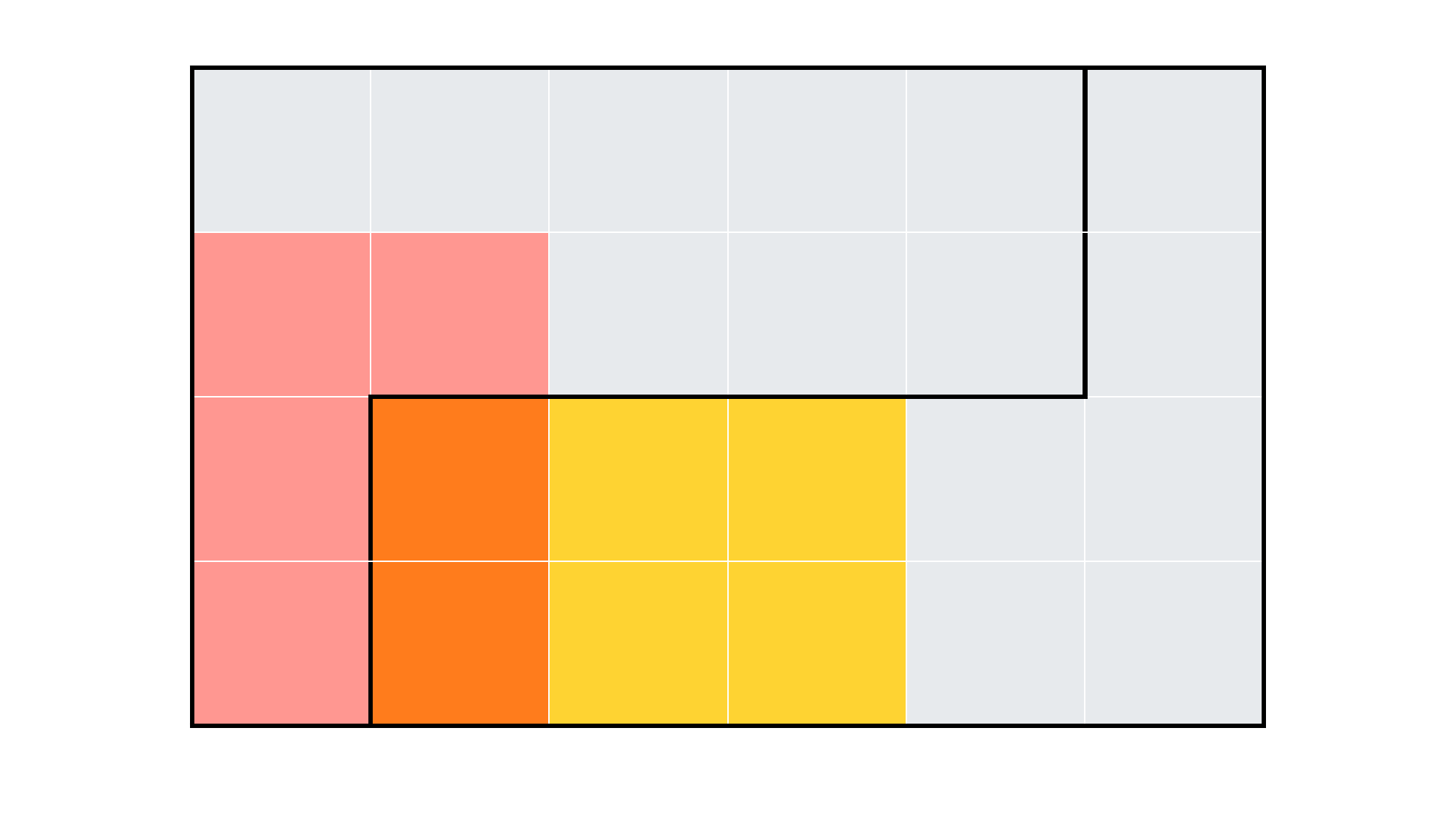}}
    \caption{\textbf{Score Functions on Gridlandia: } Two-district plans on a $4 \times 6$ Gridlandia, with two COI testimonies (pink and yellow) forming a single cluster. Panels (a) and (b) show how each score function rewards and penalizes (respectively) split COIs. Panels (c) and (d) show plans with the same testimony score and different cluster scores. }
    \label{fig:gridlandia}
\end{figure}

\medskip

In Figure \ref{fig:gridlandia}, we show some representative examples of how you calculate each score, and how they might differ for different redistricting plans. In these examples we are drawing two district plans on a $4 \times 6$ Gridlandia (coined by MGGG\footnote{\url{https://mggg.org/metagraph/}}). Possible testimonies are shown in pink and yellow, and they aggregate into Gridlandia's only COI cluster. The redistricting plan $\xi_a$ shown in Figure \ref{fig:gridlandia} (a) shows a plan that keeps all testimonies whole, so $s_t(\textbf{t}, \xi_a) = s_c(\textbf{t}, \xi_a) = 1$. Figure \ref{fig:gridlandia} (b) shows a plan that splits all testimonies, so $s_t(\textbf{t}, \xi_b) = 0$. To calculate the cluster score, note $w^1 = w^2 = 6$ (we omit $j$ as there is only one cluster) and $W = 12$, thus $s_c(\textbf{t}, \xi_b) = \frac{1}{2}$, the minimum cluster score for one cluster and two districts. Panels (c) and (d) of the figure show how the cluster score can be more expressive than the testimony score. Note that $s_t(\textbf{t}, \xi_c) = s_t(\textbf{t}, \xi_d) = \frac{1}{2}$ even though the district line in (d) cuts out most of the pink testimony. This is reflected in the cluster score where $s_c(\textbf{t}, \xi_c) = \frac{11}{12}$ while $s_c(\textbf{t}, \xi_d) = \frac{8}{12}$. 

\medskip

\begin{figure}[!htbp]
    \centering
    \includegraphics[width=\linewidth]{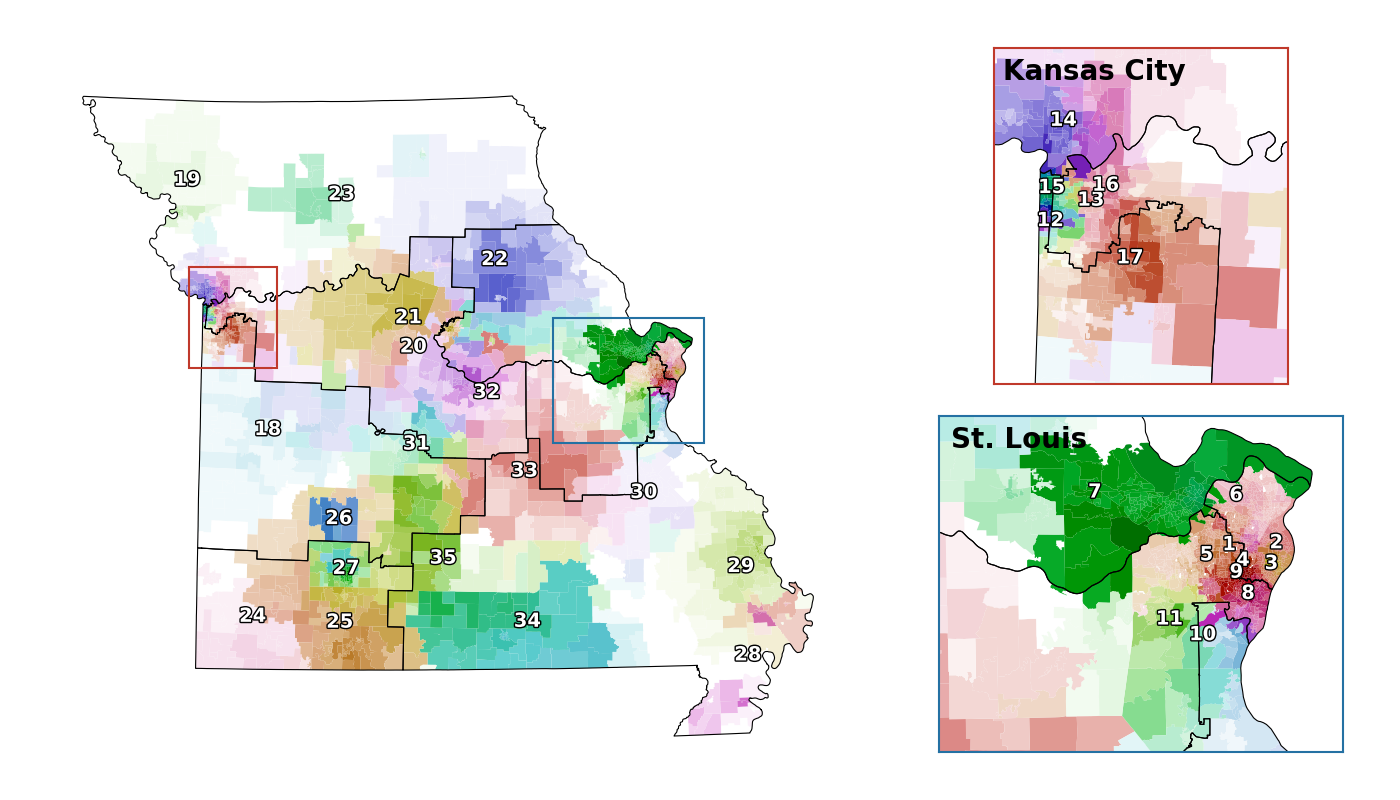}
    \caption{\textbf{COI Clusters and Enacted Plan: } The 2025 enacted plan is shown over the COI clusters calculated in \cite{mggg_missouri_coi}. Insets show detailed maps of St. Louis and Kansas City. In the latter, the district line cuts through the heart of the city, and through clusters 12-16.}
    \label{fig:enacted}
\end{figure}

In Figure \ref{fig:enacted}, we plot Missouri's enacted plan over the clusters found by \cite{mggg_missouri_coi}. The enacted plan has $s_t(\textbf{t}, \xi)= 0.62$ and $s_c(\textbf{t}, \xi)=0.71$. 

\medskip

With our two score functions, we define our two target distributions. Since we want to sample valid and realistic redistricting plans, we add a compactness term to the target distribution in the form of a gaussian with respect to cut edges centered on the enacted plan's value and with a standard deviation of 10 edges. Our target distributions take the form: $$\pi(\xi) \propto e^{-(J_{\text{compactness}}(\xi) + J_{\text{privacy}}(\xi))}$$ where $J_{\text{compactness}}(\xi) =  -\beta \cdot (c(\xi)-\mu)^2$ is a normal distribution over cut edges $c$ with mean $\mu$ and variance $\frac{1}{2\beta}$. The enacted plan in Missouri has 631 cut edges. Thus, in order for the ensemble to be normally distributed around 631 with a standard deviation of 10, we choose $\mu = 614, \beta=0.02$ to account for the shift observed in \cite{mcwhorter2025marked}. We let $J_{\text{privacy}}$ be the exponential mechanism, i.e. $J_{\text{privacy}} = \frac{\epsilon}{2\Delta(s_i)} \cdot s_i(\textbf{t}, \xi)$, where $s_i$ can be either $s_t$ or $s_c$.

\subsection{Mixing Heuristics}





\begin{figure}[!htbp]
    \centering
    \includegraphics[width=0.5\linewidth]{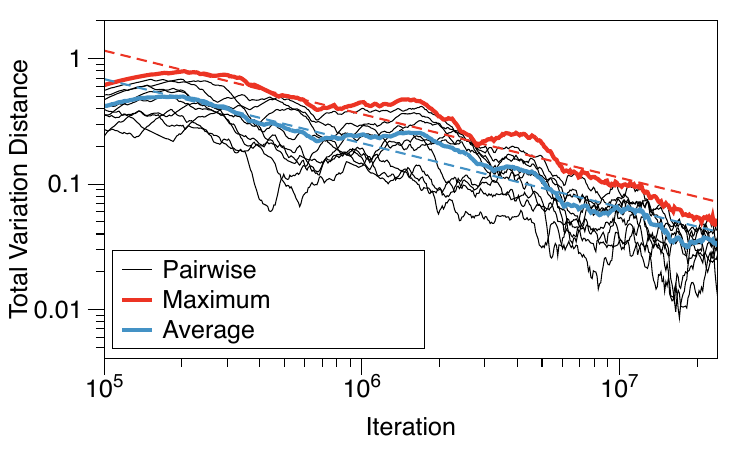}
    \caption{\textbf{Total Variation Distance Decay:} We plot the maximum (red), average (blue), and each pairwise (black) TVD vs. iteration. The maximum and average decay rates approximately follow a power law of $O(n^{-0.51})$. }
    \label{fig:TVD_decay}
\end{figure}

We assess convergence by running five independent chains for each combination of score function and privacy budget, each initialized from a random seed drawn by the bipartition tree algorithm included in gerrychain \cite{DeFord2021Recombination}. To set the length of the chains, we calculate pairwise total variation between the empirical score distributions of each chain, and set as a stopping criterion that the average pairwise TVD should be less than 0.05. An example decay of this value for the testimony score with $\epsilon=0$ is shown in Figure \ref{fig:TVD_decay}, and the decay follows a power law of $O(n^{-0.51})$. Similar power law behavior is observed for both score functions and across privacy budgets within the usable range. Beyond a threshold privacy budget, the chains do not mix. Likely, this is due to the fact that when $\epsilon$ increases, the exponential mechanism assigns vanishingly small probabilities to proposals that decrease the score function, trapping the chain. We exclude these cases from the analysis. The usable ranges are $\epsilon<1$ for the testimony score and $\epsilon<100$ for the cluster score. The lower threshold for the testimony score is likely due to the fact that $s_t$ changes in discrete steps of $\frac{1}{T}$, while $s_c$ can change by smaller amounts, giving the chain more maneuverability at higher budgets. 

\medskip

As noted in Section \ref{sec:DP}, whenever the target distribution is $(\epsilon, 0)$-DP and the Markov chain converges uniformly, sampling from the empirical distribution is $(\epsilon, \delta)$-DP with $\delta = R(n)(e^\epsilon + 1)$, where $R(n) = \max_{\nu \in \mathcal{P}(X)}||\nu P^n-\pi||_{TV}$ is the decay-rate of the worst-case total variation distance from stationary, and $\mathcal{P}(X)$ denotes the set of probability distributions over $X$. By the triangle inequality, the worst-case TVD between two initial distributions $\bar{R}(n)=\max_{\mu, \nu \in \mathcal{P}(X)}||\mu P^n - \nu P^n||_{TV} \geq R(n)$. Thus, as a heuristic for $R(n)$, we calculate the maximum pairwise TVD between empirical score distributions. This heuristic is likely an underestimate of $R(n)$ since we have a finite set of initial states rather than the full distribution space. The requirement that $\delta < 1$ translates to $R(n)(e^\epsilon + 1) < 1$. At $\epsilon \leq 1$ this requires $R(n) \leq (e+1)^{-1} \approx 0.27$, which is comfortably satisfied. At $\epsilon=10$, $R(n) \leq (e^{10} + 1)^{-1} \sim O(10^{-5})$, which is not achievable in a practicable number of steps. For all cases, we report the maximum pairwise TVD when the stopping criterion is met and the estimated value for $\delta$ in Tables \ref{tab:testimony_TVDs} and \ref{tab:cluster_TVDs} of the Appendix. 

\medskip

While formal convergence guaranties are limited for algorithms of this form, empirical results have demonstrated convergence to the correct stationary distribution on small graphs and to Gaussian distributions over cut edges on larger graphs \cite{mcwhorter2025marked}. Combined with the mixing diagnostics above, this gives reasonable confidence that our chains are sampling from the target distributions. 

\section{Results}
\label{sec:results}

\subsection{Score Distributions at Differing Privacy Budgets}

\begin{figure}[!htbp]
    \centering
    \subfigure[]{\includegraphics[width=0.49\textwidth]{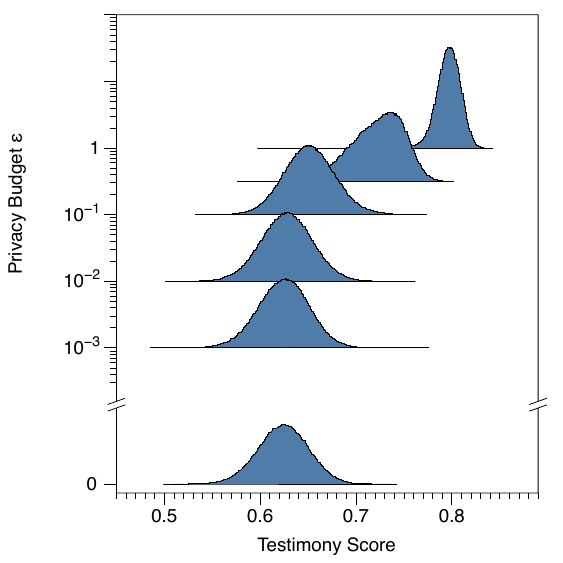}}
    \subfigure[]{\includegraphics[width=0.49\textwidth]{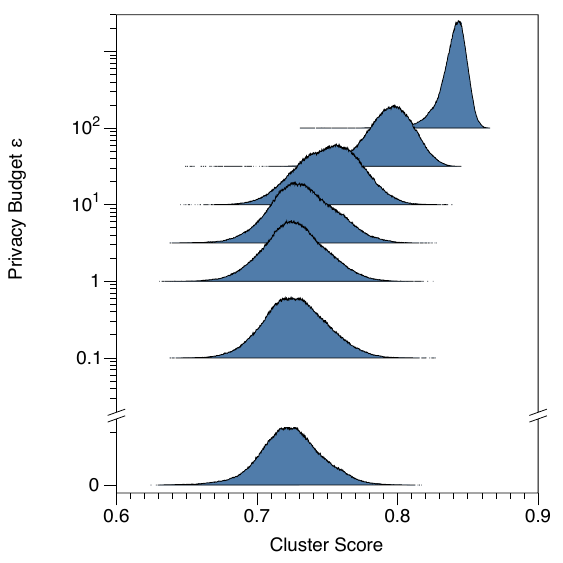}}
    \caption{\textbf{Score Distributions across Privacy Budgets:} Score distributions of the testimony score (a) and cluster score (b) across different privacy budgets. As $\epsilon$ increases, the expected loss when sampling from the distributions decreases, illustrating the privacy-loss trade off. }
    \label{fig:ridges}
\end{figure}

Figure \ref{fig:ridges} shows the sampled distribution of each score function at increasing privacy budgets. Both score functions display similar behavior: as $\epsilon$ increases, the distributions shift towards higher scores and become narrower, indicating that larger privacy budgets produce plans that better preserve COIs. 
\medskip

An interesting result of both figures is the baseline distribution at $\epsilon=0$. Even without any COI weighting, plans sampled from the compactness distribution preserve on average over 62\% of testimonies and over 72\% of cluster mass. These are surprisingly high baseline scores. For context, the enacted plan has $s_t(\textbf{t}, \xi_{\text{enacted}}) = 0.62, s_c(\textbf{c}, \xi_{\text{enacted}}) = 0.71$. It could be that the compactness requirement incidentally keeps many geographically compact communities together. Moreover, most testimonies are small relative to the size of a district, making accidental preservation more likely. However, the enacted plan (Figure \ref{fig:enacted}) does a noticeably poor job at preserving COIs, so it is clear we still need to improve on this baseline. At low privacy budgets, the distributions are very similar to the baseline, suggesting the compactness term dominates the exponential mechanism in this regime. A shift to higher COI scores becomes visible only at moderate to high $\epsilon$, signaling the COI weighting is beginning to make an impact.



\subsection{Score Comparison}





Given the relationship between testimonies and clusters, it is natural to suspect there may be a relationship between the testimony score and the cluster score. To assess this, we calculate both scores on every sampled plan regardless of which score function governed the chain. That is, plans sampled under $s_t$ are also evaluated on $s_c$, and vice versa. Based on the first fifteen million iterations of five independent runs, we find moderate positive correlations between the two scores across all $\epsilon$ values and both score functions, ranging from $r =0.405$ to $r = 0.532$ (see Table \ref{tab:cors} of the Appendix). Under $s_c$, the correlation shows no clear trend with $\epsilon$, while under $s_t$, the correlation increases as the privacy budget decreases. This is consistent with the intuition that at high $\epsilon$, the testimony score pushes plans toward keeping testimonies whole, which in turn concentrates cluster weight as well, whereas optimizing for $s_c$ might involve prioritizing high-weight cluster centers at the expense of block groups with low weight, leading to split testimonies. 

\medskip

In both cases, the cluster score is almost always larger than the testimony score for a given plan, even when the testimony score governs the chain. However, the governing score dominates more as $\epsilon$ increases; the least-private testimony chain ($\epsilon = 1)$ is the only instance where more than 75\% of the plans have a larger testimony score. And, in the cluster score governed chains, the highest three privacy budgets have no plan with a higher testimony score. The difference $s_t - s_c$ ranges from -0.247 to 0.07. This tendency for $s_c > s_t$ reflects the finer-grained nature of the cluster score: a small divide of a large COI may drop its testimony score contribution to zero, while its cluster score contribution falls only fractionally. For full statistics on the score differences, see Tables \ref{tab:s-t for c} and \ref{tab:s-c for t} of the Appendix.

\subsection{Preservation of Different COIs}

Testimony split frequency is correlated with several natural properties of the testimonies, including population, area, and structural graph properties such as the relative number of internal to boundary nodes, the relative number of internal to boundary edges, and spanning trees. However, while these correlations are statistically detectable, they are moderate, and come with high residual variance. Population has the strongest correlation with split frequency at $\epsilon = 0$ ($r\approx 0.523$), followed by the logarithm of spanning trees ($r \approx 0.520$), while relative internal edges, nodes, and area show weaker relationships (see Table \ref{tab:cor_prop} and Figures \ref{fig:area_split} and \ref{fig:pop_split} of the Appendix). The limited predictive power of these relationships makes it challenging for a bad actor to design a testimony that is reliably preserved based on these properties alone. The geographic location of a testimony and its relationship to the surrounding plan space matter at least as much as any single measurable property. 

\medskip

\begin{figure}[!htbp]
    \centering
    \subfigure[]{\includegraphics[width=0.42\linewidth]{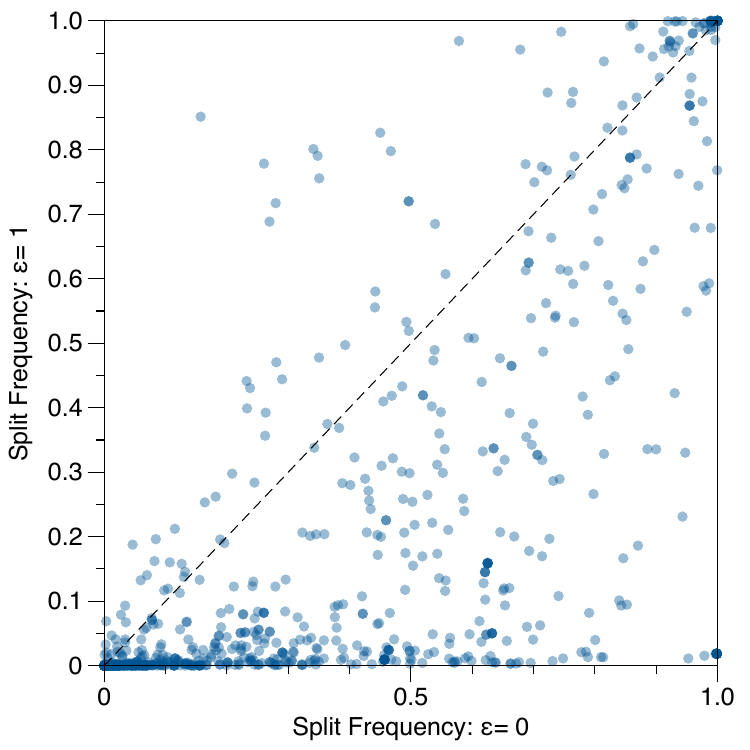}}
    \subfigure[]{\includegraphics[width=0.42\linewidth]{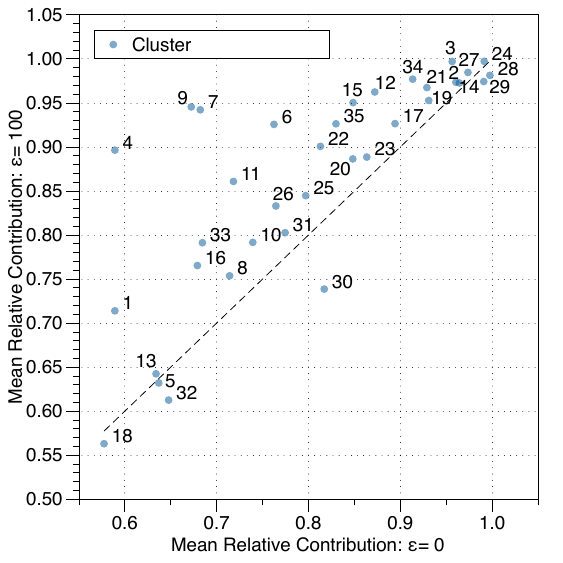}}
    \caption{\textbf{Score Contributions at Different Privacy Budgets: } Scattered testimonies' split frequencies (a) and relative cluster score contribution (b) at the largest and smallest privacy budgets. In panel (a), testimonies that lie above $y = x$ are split more as the privacy budget increases, indicating that the algorithm may `sacrifice' some testimonies to prioritize others. In panel (b), clusters generally contribute more to the cluster score at a higher privacy budget. }
    \label{fig:split_split}
\end{figure}

A more direct consequence of COI-informed sampling is the sacrifice phenomenon. As the privacy budget increases, some testimonies are split at \textit{higher} rates so that others can be kept whole. Panel (a) of Figure \ref{fig:split_split} shows which testimonies are affected. Overall, 16.1\% of testimonies are split more often at $\epsilon = 1$ than at $\epsilon = 0$, with the worst instance increasing from a 15.7\% split rate to 85.1\%. This is a consequence of the all-or-nothing nature of the testimony score. Once the exponential mechanism begins to dominate, the sampler optimizes globally, and some testimonies in critical locations are traded off to preserve others. Panel (b) of Figure \ref{fig:split_split} shows that clusters exhibit similar but less pronounced behavior. Cluster 30, the smallest and least dense cluster, shows the most sacrifice, while cluster 18 consistently has the lowest mean relative contribution across all epsilon values, as is expected given its sprawling geography and few testimonies. Clusters in and around St. Louis (4, 6, 7, and 9) improve significantly as privacy decreases. Cluster 24 maintains the highest retention rate across all epsilon values, likely due to its location in the southeast corner of the state.

\medskip

At the extremes, split frequency is largely determined by population and size. There are eighteen testimonies that are always split regardless of epsilon, located in clusters 1, 5, 13, and 25. Clusters 1 and 13 carry a large weight in these oversized COIs, which is reflected in their poor relative contribution to the cluster score. At the other extreme, nineteen testimonies are never split, fourteen of which consist of a single block group.

\subsection{Adversarial Testimony Experiment}







\begin{figure}[!htbp]
    \centering
    \subfigure[]{\includegraphics[width=0.49\linewidth]{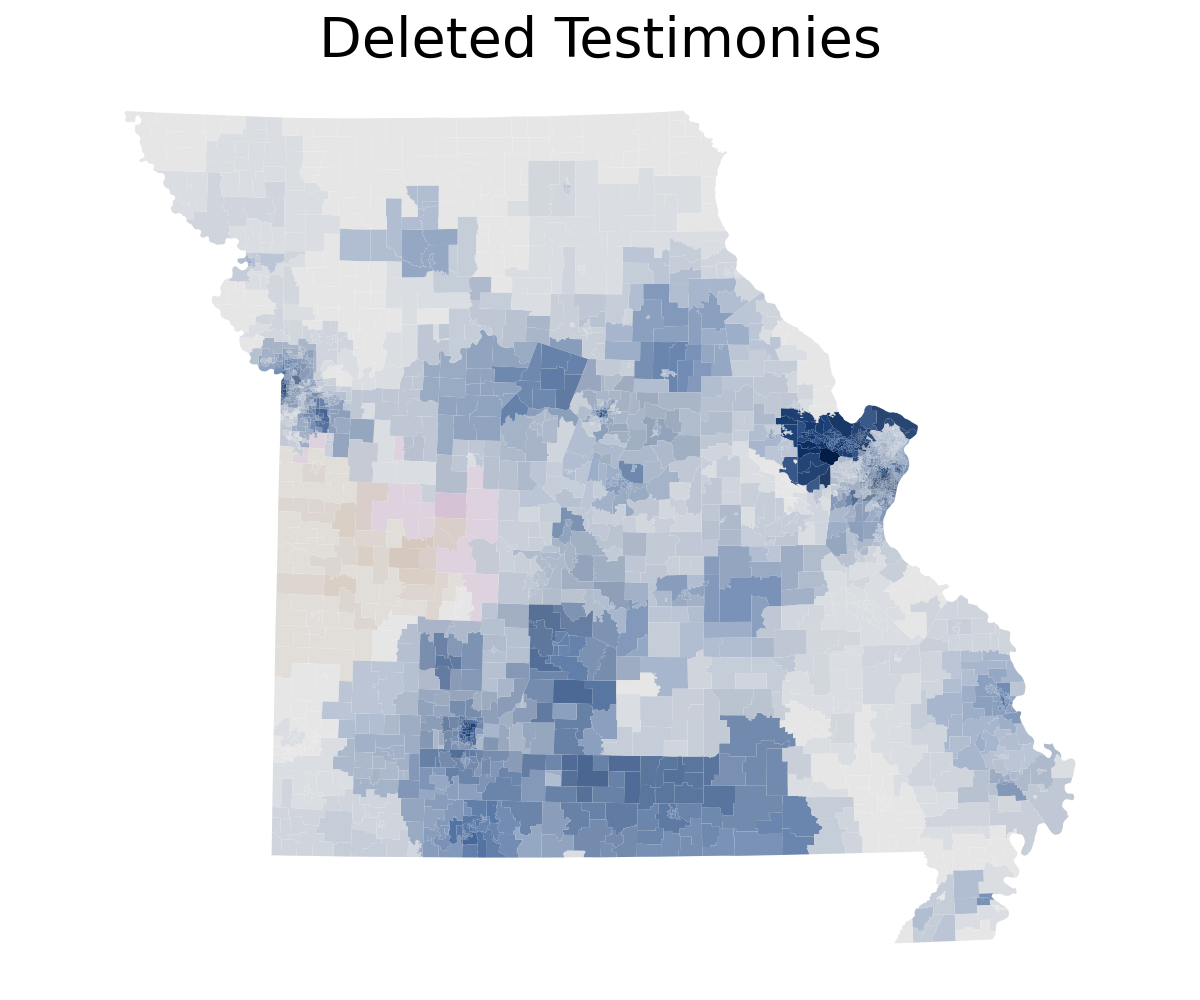}}
    \subfigure[]{\includegraphics[width=0.49\linewidth]{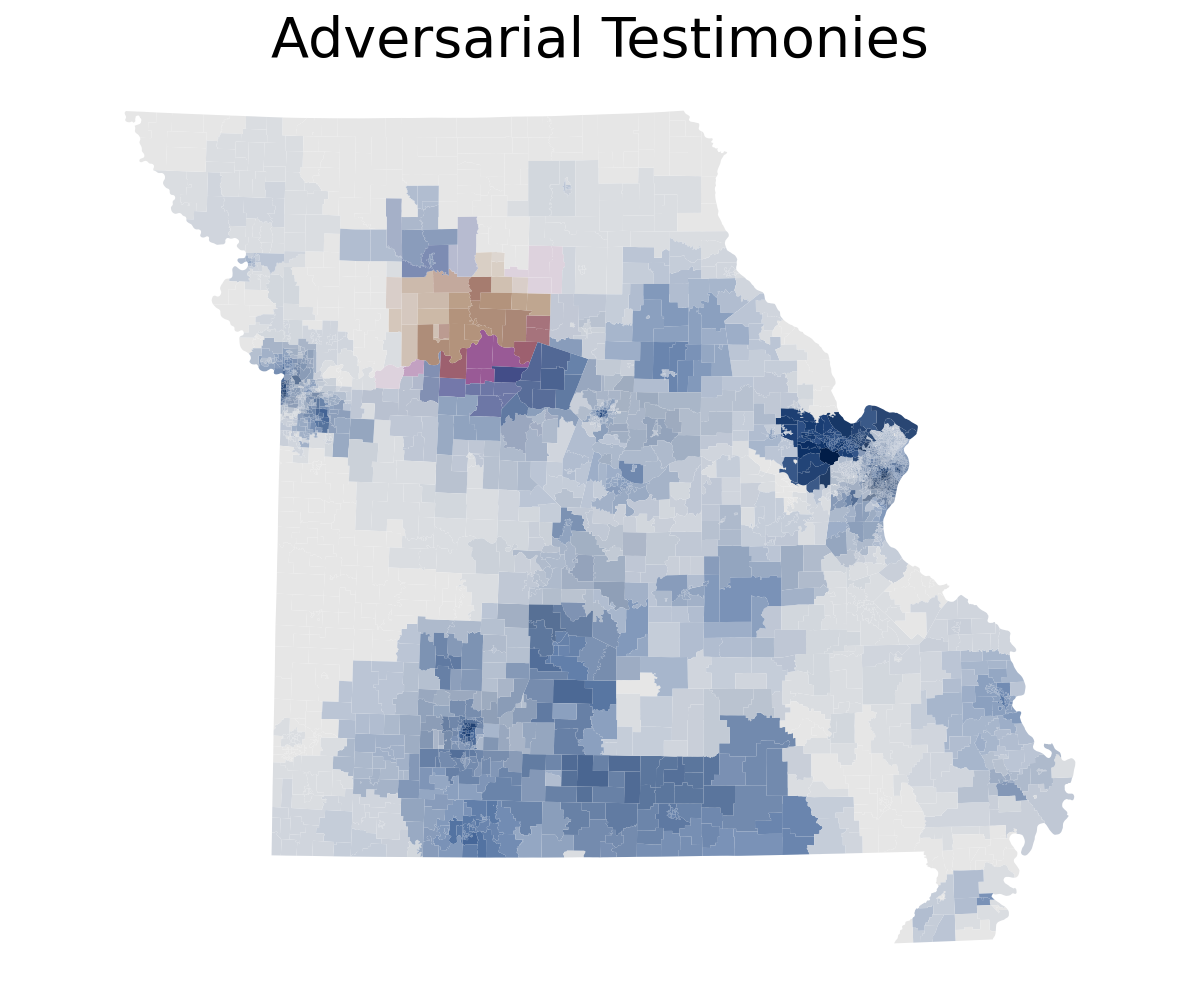}}
    \caption{\textbf{Adversarial Testimony Comparison: } Actual (a) and Adversarial (b) COI clusters, with differences highlighted in yellow. The adversarial experiment compares probabilities of redistricting configurations under the two testimony sets.}
    \label{fig:adv_map}
\end{figure}

As described in Section \ref{sec:intro}, coordinated manipulation of IRC testimony gathering has occurred in practice. In this section, we describe an experiment in which we introduce adversarial testimonies into the dataset and analyze their effect on the sampled distributions. Figure \ref{fig:adv_map} shows the original and adversarial data sets, with differences highlighted in yellow. All nine testimonies of cluster 18 were removed and replaced with nine adversarial testimonies covering a geographically distinct area. We then run the chain on the adversarial dataset across a range of privacy budgets and both score functions, and compare the outputs to those obtained with the original dataset. 

\medskip

\begin{figure}[!htbp]
    \centering
    \subfigure[]{\includegraphics[width=0.49\linewidth]{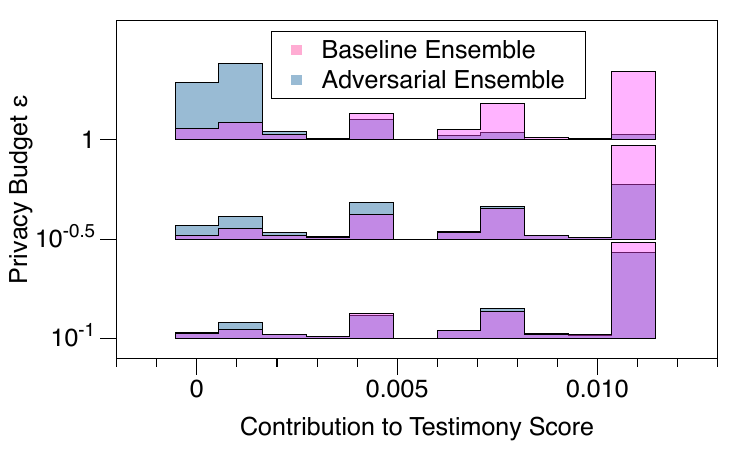}}
    \subfigure[]{\includegraphics[width=0.49\linewidth]{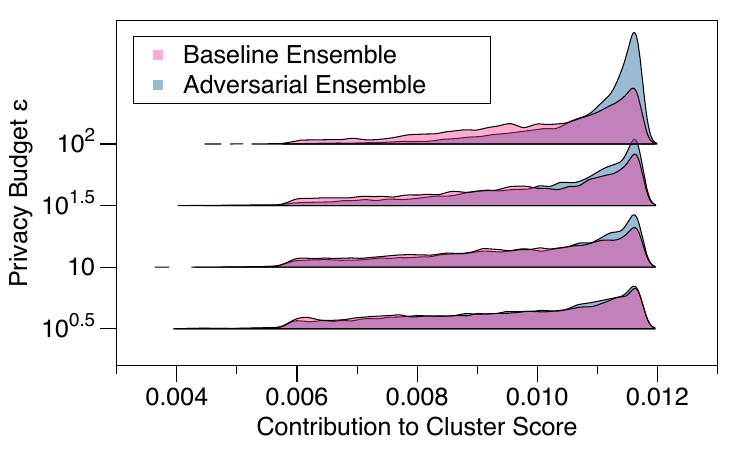}}
    \caption{\textbf{Score Contributions of Adversarial Testimonies: } Testimony (a) and cluster (b) score contributions of the adversarial testimonies across privacy budgets over ensembles of the baseline data (blue) and adversarial data (pink). At low values of $\epsilon$, there is little qualitative difference between the distributions. As $\epsilon$ increases, the adversarial testimonies are split more frequently by the testimony score, and kept together more frequently by the cluster score. }
    \label{fig:adv_contrib}
\end{figure}

Figure \ref{fig:adv_contrib} shows the contribution of adversarial testimonies to each score function under baseline plans, sampled using the original testimonies, and adversarial plans, sampled using the adversarial data. Pink distributions show the adversarial testimonies' score contributions under the baseline plans, and blue distributions show their contributions under the adversarial plans. Since the baseline sampler does not see the adversarial testimonies, the pink distributions are unaffected by $\epsilon$. The blue distributions tell a different story. Under the cluster score (panel (b)), the adversarial cluster is held together at roughly the baseline rate at low $\epsilon$, and as the privacy budget increases, the plans concentrate more weight in the adversarial cluster, consistent with the expected behavior of the exponential mechanism. Under the testimony score (panel (a)), the behavior is reversed; as $\epsilon$ increases, the adversarial testimonies are split at higher rates, sacrificed by the sampler in favor of preserving other testimonies. This is consistent with the sacrifice phenomenon described in the previous section. Note that panel (a) is a histogram rather than a density because the testimony score changes in discrete steps of $\frac{1}{T}$. 

\medskip

\begin{figure}[!htbp]
    \centering
    \includegraphics[width=0.5\linewidth]{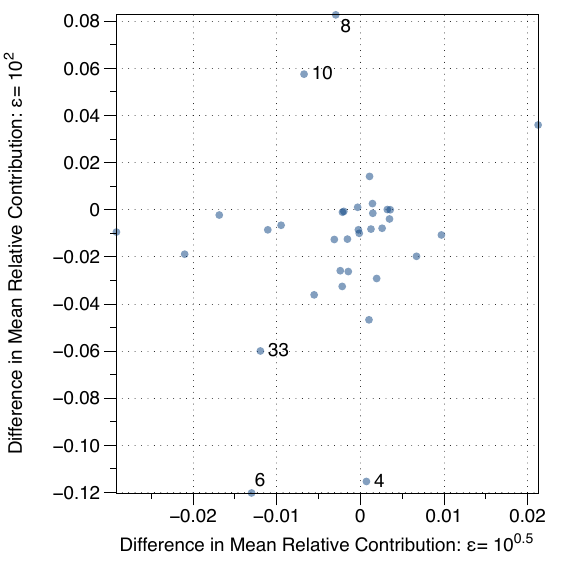}
    \caption{\textbf{Relative Score Differences of Other Clusters:} Comparison of the difference of score contributions for every cluster except 18 at a low privacy budget and a high privacy budget. At a lower privacy budget, the variance is much lower, indicating that the adversarial cluster is having a lesser global effect.}
    \label{fig:diffs}
\end{figure}

Figure \ref{fig:diffs} shows how the adversarial cluster affects the individual retention of the remaining clusters. We compare the difference in mean relative cluster score contributions between the adversarial and original datasets at two privacy levels, $\epsilon = 10^{0.5}$ and $\epsilon = 10^3$. The variance is considerably higher at higher epsilon, driven by a small number of clusters with significant changes. Three clusters are notably disfavored (clusters 4, 6, and 33) and two are bolstered (clusters 8 and 10). With the exception of cluster 33, all of these clusters are located in the St. Louis area, having no direct geographic connection to the adversarial cluster. The adversarial cluster does overlap with cluster 20 - 23, none of which show significant changes, suggesting that proximity is not enough to assess the downstream affects of adversarial attacks. 

\subsection{Partisan and Demographic Analysis}
\label{sec:partisan}

\begin{figure}[!htbp]
    \centering
    \subfigure[]{\includegraphics[width=0.49\linewidth]{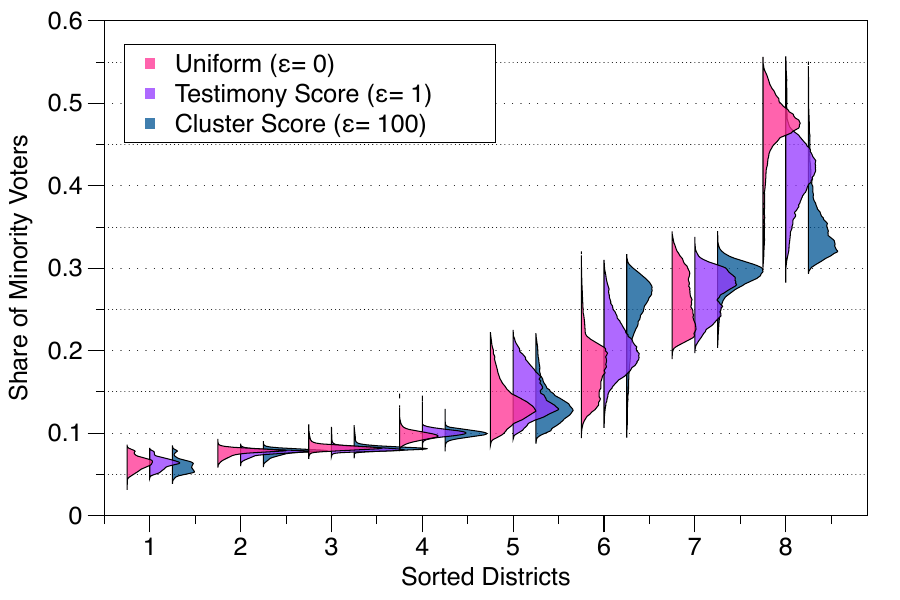}}
    \subfigure[]{\includegraphics[width=0.49\linewidth]{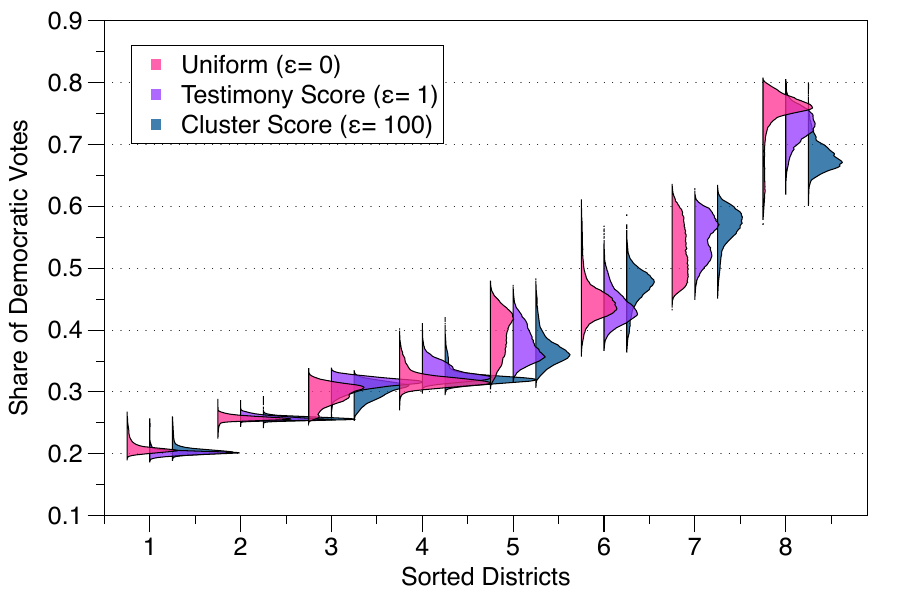}}
    \caption{\textbf{Sorted Demographic and Partisan Districts at Different Score Functions:} Sorted district distributions of MVAP proportion (a) and Democratic vote share (b). The score functions moves the mass from the highest MVAP/Democratic district across the top three districts compared to the baseline, the cluster score to a greater extent.}
    \label{fig:sorted}
\end{figure}

Figure \ref{fig:sorted} shows the distributions of sorted minority voting age population (MVAP, a) and Democratic vote share (b) by district at representative privacy budgets. In pink, we plot the baseline distribution with $\epsilon = 0$, and in blue and purple, we plot the least private distributions of the two score functions ($\epsilon = 1$ and $\epsilon = 100$, respectively). Both panels display a somewhat counterintuitive pattern: rather than concentrating minority population or Democratic votes into a single packed district, higher privacy budgets tend to spread representation more evenly across highest three ranked districts. 

\medskip

In the first panel, Missouri's demographics make it unlikely that a valid redistricting plan can produce a majority-minority district. The enacted plan includes a district with a 45\% share of Black voting age population (BVAP). In the absence of a majority-minority district, the relevant threshold becomes minority influence districts, defined as districts which have a 25\% - 40\% MVAP. At high $\epsilon$, especially under the cluster score, the weight of the sixth and seventh ranked districts moves into this range. This is consistent with the intuition that preserving COIs tends to protect minority communities, resulting in this case in more minority influence districts rather than a single packed district. 

\medskip

In the second panel, the key threshold is 50\%, at which the party wins a seat. The most notable effect of increased $\epsilon$ here is that the seventh ranked district moves from straddling the 50\% threshold to sitting solidly above it, suggesting that COI preservation tends to produce a more reliably Democratic seat. We see a similar pattern as before, where the coarse-level intuition that preserving COIs leads to packing a single Democratic district is incorrect. Instead, a higher $\epsilon$ is associated with a lower Democratic share in the highest ranked district, and gains distributed across other districts. One explanation for this (which can also be applied to panel (a)) is that Missouri's two major cities, Kansas City and St. Louis, are two distinct Democratic (and minority) communities. Combining them into a single packed district is geographically unlikely (if not impossible). So keeping COIs together actually makes it more likely that Kansas City's district is preserved enough to win a seat. We note that the enacted plan splits Kansas City in a manner that dilutes its Democratic (and minority) representation, as shown in Figure \ref{fig:enacted}. The low ranked districts show little change between the baseline and COI-informed distributions in either panel, which may reflect that most COIs are located in and around urban areas, so changes in how they are split have little effect on the rural districts. 

\medskip

\begin{figure}[!htbp]
    \centering
    \subfigure[]{\includegraphics[width=0.42\textwidth]{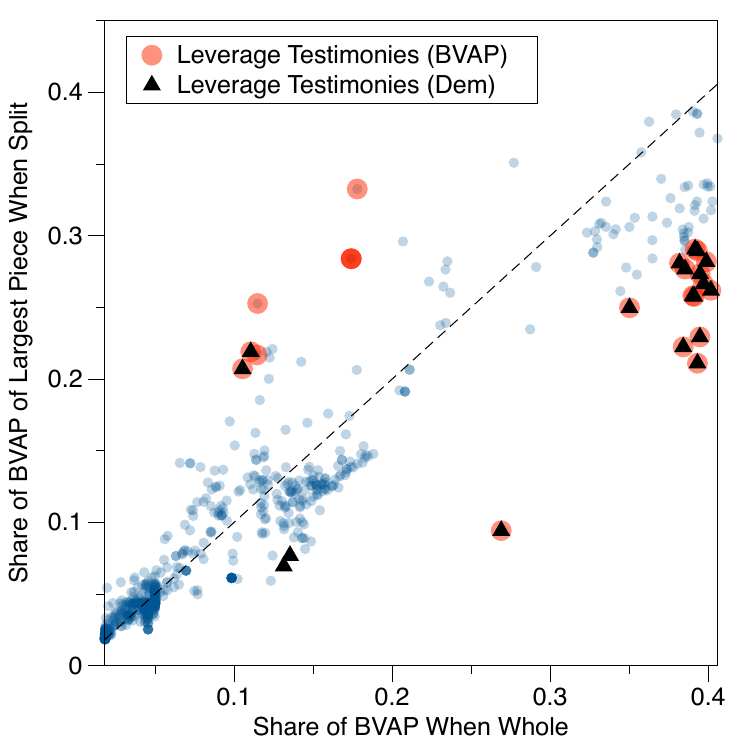}}
    \subfigure[]{\includegraphics[width=0.42\textwidth]{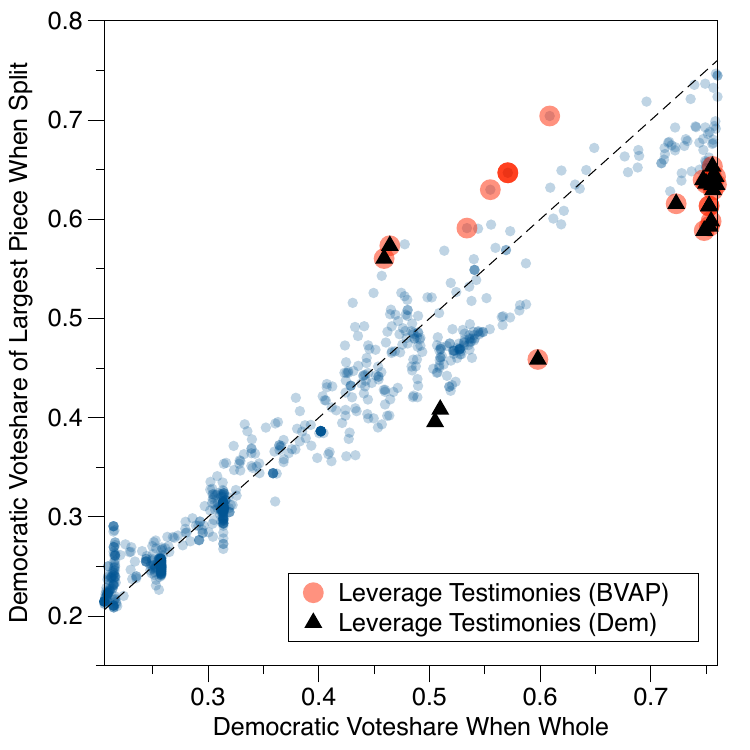}}
    \subfigure[]{\includegraphics[width=0.42\textwidth]{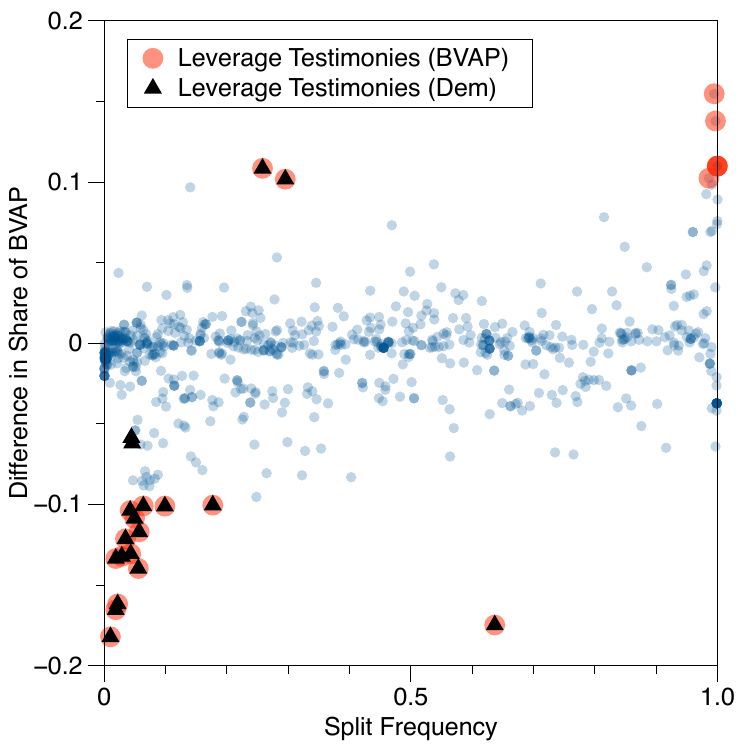}}
    \subfigure[]{\includegraphics[width=0.42\textwidth]{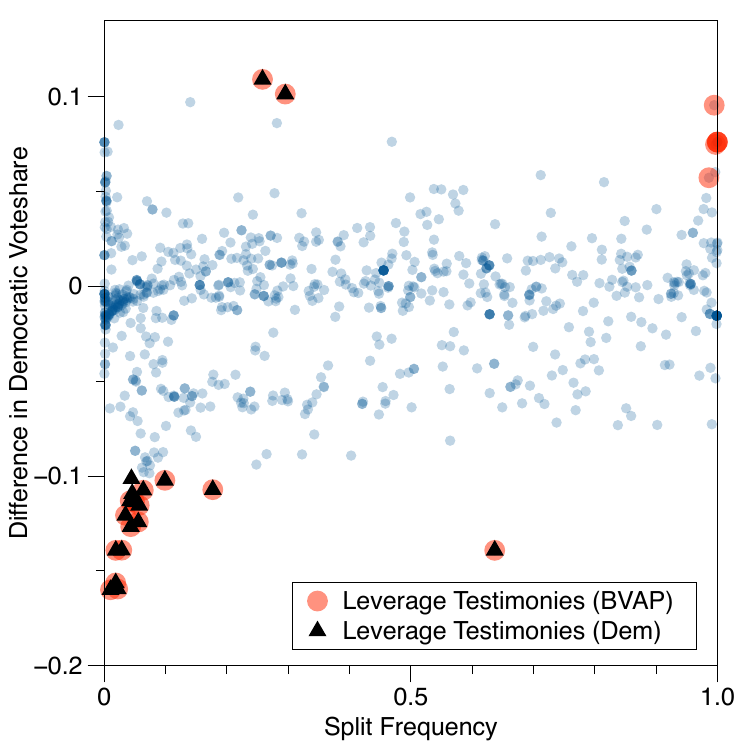}}
    \subfigure[]{\includegraphics[width=0.42\textwidth]{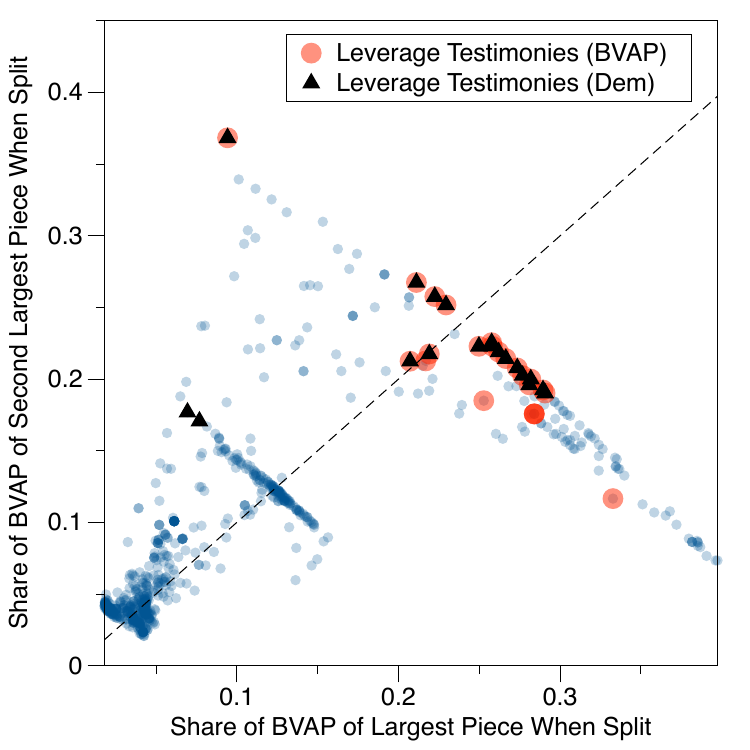}}
    \subfigure[]{\includegraphics[width=0.42\textwidth]{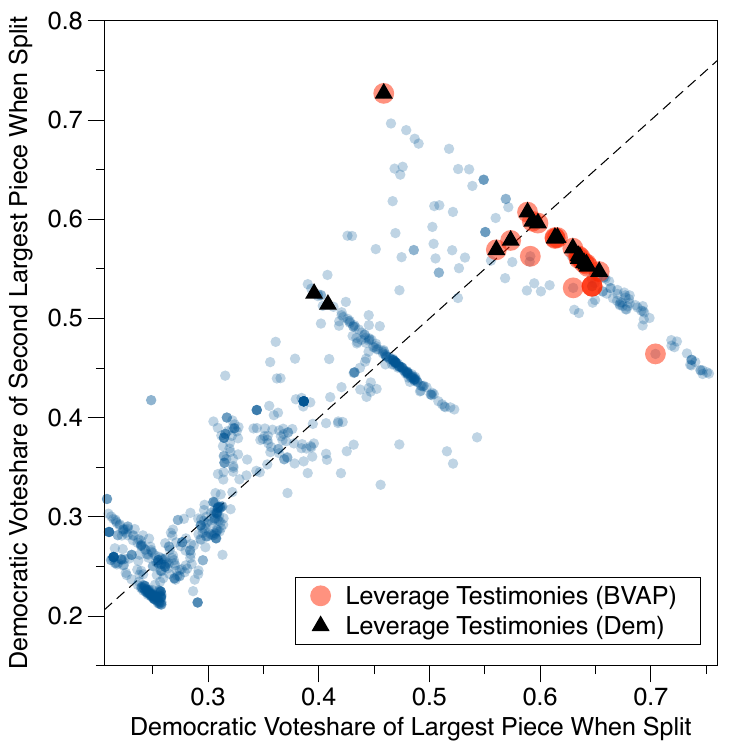}}
    \caption{\textbf{Leverage COI Investigation: } Scatter plots that identify leverage COIs with respect to BVAP (a) and Democratic vote share (b), investigate their split rates (c, d), and compare the racial (e) and partisan (f) makeup of their parts when split. }
    \label{fig:mega_leverage}
\end{figure}

We now investigate leverage COIs, testimonies that have a meaningful effect on the partisan or demographic character of the resulting districts. We say a testimony is a leverage testimony if the Democratic vote share or BVAP of the district containing its largest piece when split differs from the value when whole by at least 10\%. This analysis shows little dependence on $\epsilon$, and thus we average over all usable privacy budgets of the testimony score. Figures \ref{fig:mega_leverage}(a) and (b) plot the split share against the whole share for BVAP and Democratic vote share respectively, with leverage testimonies highlighted in red (BVAP) and black triangles (Democratic vote share). Most testimonies cluster near the $y=x$ line, indicating that splitting has little effect on their demographic or partisan makeup. Leverage testimonies show up on either side of this line. Those below the line have a lower share when split, meaning their constituency is diluted when split, while those above the line have a higher share in their largest piece when split. It is also of note that many of the leverage testimonies coincide;. a testimony that is a leverage testimony with respect to BVAP is often a leverage testimony with respect to Democratic vote share.

\medskip

Panels (c) and (d) plot the difference in share (whole minus largest piece when split) against split frequency. Testimonies that are diluted when split (that lie below the $y=x$ line in panels (a) and (b)) tend to have low split frequencies, while those that are split most frequently tend to have a positive difference. This suggests that an implicit effect of our method is that the plans sampled tend to protect the testimonies whose splitting would cause the most dilution, although we have not formally tested this proposition. 

\medskip

Panels (e) and (f) examine the second largest piece when split, plotting its share against the share of the largest piece. This looks at a different idea of leverage; rather than asking about the global partisan or demographic effect of splitting a COI, we investigate whether splitting divides a constituency into two districts with very different characteristics. In these panels, most of the previously defined leverage testimonies cluster near the $y=x$ line, suggesting that while they have meaningful global effects when split, their constituencies are not (typically) divided into two districts with differing racial or partisan makeup. One outlier to this pattern is visible in both panels. 

\medskip


\begin{figure}[!htbp]
    \centering
    \includegraphics[width=0.7\linewidth]{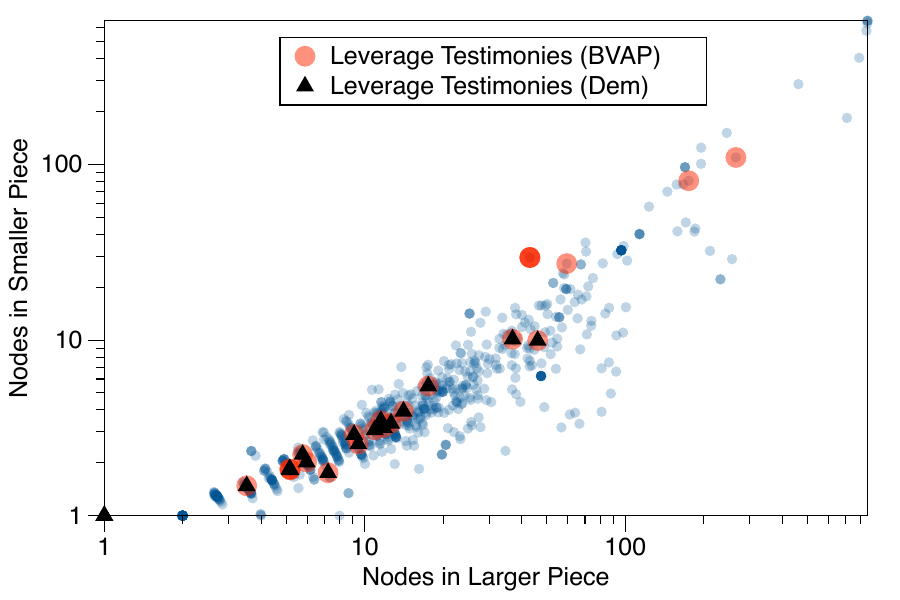}
    \caption{\textbf{Piece Sizes of Split Testimonies: } Relative size of the largest and second largest piece when testimonies are split. Leverage testimonies fall in the middle of the distribution, indicating there is no relationship between the splitting pattern of a testimony and whether it is a leverage testimony. }
    \label{fig:large_small}
\end{figure}

Finally, Figure \ref{fig:large_small} plots the size of the second largest piece against the size of the largest piece for split testimonies, showing that the leverage testimonies are distributed similarly to non-leverage testimonies. This confirms that the leverage effects are not results of lopsided splits in which a single node is separated from the rest of the testimony.

\section{Discussion}


\paragraph{Motivation} 
IRCs represent a promising solution for bottom-up redistricting, but as we noted in the introduction, their testimony processes are vulnerable to adversarial manipulation. We use differential privacy to draw plans that can take public input while remaining robust to adversarial testimonies. In our application, we treat individual testimonies as data points and use MEW to sample from differentially private distributions of redistricting plans. This ensures that the probability of any given plan under two neighboring testimony datasets cannot differ by more than a quantifiable bound. Using the exponential mechanism, we show that MEW (or more generally, any sampler that can be Metropolized to target distributions) can be used to sample heuristically private redistricting plans. Across differing privacy budgets and score functions, we show it is possible to sample plans that are private to the testimony data while still outperforming an uninformed baseline (and thus, the enacted plan). 

\medskip

\paragraph{General Behavior}
Before discussing the private sampling, it is worth noting that the baseline distributions still perform relatively well. The compactness term ensures that sampled plans have similar compactness to the enacted Missouri districts, and even under this uninformed distribution, the enacted plan falls on the worse end of the score distribution. This says something about the enacted plan itself, which splits Kansas City in a way that dilutes the Black population across three districts. And more generally, a seemingly naive compactness constraint can produce fine COI preservation scores purely due to geography. People tend to propose geographically coherent, blob-like testimonies, which fit naturally into compact districts. 

\medskip

Despite the baseline performing well, there remains meaningful room for improvement through COI-informed sampling. As expected, expected loss decreases as the privacy budget increases (as we prioritize privacy less and COI preservation more). This raises the classic privacy-utility tradeoff, though we are more constrained than many DP applications. At low $\epsilon$, the compactness term dominates the exponential mechanism and the distributions are not meaningfully different from the baseline. At high $\epsilon$, the exponential mechanism concentrates and the probability of accepting score-decreasing proposals becomes too small for the chain to mix well. This limits us to a practicable range of privacy budgets.

\medskip

\paragraph{Score Functions}
These two score functions tell a similar story in different ways, and each has their advantages. The testimony score is easily explainable, simple to implement, and is modifiable. The testimony score also carries stronger privacy guarantees due to its smaller sensitivity. The tradeoff is some counterintuitive behavior: as the privacy budget increases, some testimonies are split at higher rates so that other can be kept whole, a consequence of the discrete, all-or-nothing nature of the score. The cluster score is in many ways the inverse. It requires an additional clustering step and is harder to explain in a political context. However, it better captures the spirit of COI preservation; individual testimonies are aggregated into a broader community, and it is the community which is scored. Because changes to the score are finer, the chain mixes better at higher privacy budgets, and the results tend to be more intuitive and pronounced. The cost is weaker privacy guarantees due to the higher sensitivity. For practitioners primarily interested in minimizing loss, the cluster score appears to be the stronger choice. 

\medskip

This difference is well exemplified by the adversarial experiment. Under the cluster score, the response is as expected. At lower $\epsilon$, the adversarial cluster is held together at roughly the baseline rate, and as the privacy budget increases, the plans keep it together more. Under the testimony score the behavior is reversed. At low $\epsilon$, the ensemble agrees with the baseline, but as $\epsilon$ increases, the adversarial testimonies are split at higher rates, sacrificed so that other testimonies can be kept whole. 

\medskip

\paragraph{Adversarial Experiment}
The adversarial experiment also suggests that the algorithm performs better in practice than the formal guarantees would imply, because the privacy guarantee protects against the worse cases. The lowest $\epsilon$ values in the experiment are $\epsilon = 0.1$ for the testimony score are $\epsilon = 10^{0.5}$ for the cluster score. With nine adversarial testimonies, the group privacy guarantee degrades to $(9\epsilon, 9\delta)$-DP, which is quite weak. Yet, at these levels, the contribution of the adversarial testimonies to the score distribution closely tracks the baseline, and the global behavior does not differ substantially. That said, the fundamental limitation of group privacy remains: a sufficiently large coordinated adversarial attack could overcome the mechanism. As the number of adversaries grows, group privacy could scale linearly, and the protection diminishes. 

\medskip

\paragraph{Additional Analysis}
The partisan and demographic analysis reveals some interesting behavior for Missouri. Due to the geographic separation of Missouri's two major minority and Democratic communities, stronger COI preservation actually unpacks the top district. In the case of MVAP, the minority population is spread more evenly across the second and third ranked districts, creating three solid minority influence districts where the baseline produces only one. The situation is parallel for Democratic vote share: COI preservation solidifies a district for the Kansas City area. Our investigation of leverage COIs shows that there are communities that, when split, can change the global partisan and demographic makeup of the map, further motivating the need for adversarial protection. Also, this finding shows that leverage alone cannot be used to identify a testimony as adversarial; the leverage testimonies identified here were gathered in earnest. 

\medskip
\paragraph{Limitations and Future Work}
This method has some limitations, many of which could be addressed by future work. MEW's theoretical guarantees are limited, forcing practitioners to rely on heuristics for mixing. This translates to reliance on heuristics for the privacy guarantees, as $\delta$ is tied to the mixing rate, which is an active area of research but is still unknown for most of the samplers currently in use. 
Practical adoption of this method faces additional hurdles. Testimony gathering is rarely done with accompanying maps, although we hope that projects like those of MGGG with Districtr will continue to become more common. 
The cluster score could be sensitive to the choice of clustering method; as discussed above there is recent research on using geographic clustering methods in this context, 
but the effects of clustering choices on the outputs of this process are unknown. Lastly, redistricting plans are not drawn from a random process in practice, and any application would require a series of consequential algorithmic choices ($\epsilon, \delta$, score function, clustering method) which involve tradeoffs that are not well understood, although these results could still be used to evaluate the properties of a plan under the ensemble framework. 

\medskip
\paragraph{Conclusion}
Redistricting from the bottom up offers the potential to return the line-drawing process to the communities that elections are meant to serve. While IRCs can  represent a significant step in this direction, the potential for adversarial input in the form of false or manipulated testimony can corrupt the IRC process. The method we present in this paper provides a novel approach towards curbing the power of adversarial testimonies while still producing maps that are informed by COIs. It does not completely solve the problem, as coordinated actors can always overpower the exponential mechanism, but it does provide a framework for making the influence of any single voice quantifiable and bounded.

\newpage

\bibliography{Gerry_refs_1}

\begin{appendices}

\section{Mixing Heuristics}

\begin{table}[!htbp]
\centering
\begin{tabular}{||l c c c c c||} 
 \hline
 Epsilon & Iterations (millions) & Average TVD & Maximum TVD & Estimated $\delta$ &\\ 
 \hline
  0 & 24 & 0.036 & 0.053 & 0.072 &   \\

 $10^{-3}$ & 35 & 0.043 & 0.068  & 0.136 &\\ 

 $10^{-2}$ & 35 & 0.057 & 0.116 & 0.233  & \\

 $10^{-1}$ & 35 &  0.047 & 0.070 & 0.147 &  \\
 
 $10^{-0.5}$ & 30 & 0.060 & 0.105 & 0.251 &  \\

 $1$ & 34 & 0.037 & 0.069 & 0.257 & \\
 \hline

\end{tabular}
    \caption{Mixing heuristics for $s_t$}
    \label{tab:testimony_TVDs}
\end{table}

\begin{table}[h]
\centering
\begin{tabular}{||l c c c c c||} 
 \hline
 Epsilon & Iterations (millions) & Average TVD & Maximum TVD & Estimated $\delta$ &\\ 
 \hline
  0         & 23 & 0.055 & 0.086 & 0.172 &   \\

 $10^{-1}$  & 20 & 0.046 & 0.063  & 0.133 & \\ 

 $1$        & 37 & 0.056 & 0.104 & 0.387  & \\

 $10^{0.5}$ & 37 & 0.038 & 0.068 & 1.674 &  \\

 $10^{1}$   & 25 & 0.044 & 0.076 & $1.7\cdot10^3$ &  \\

 $10^{1.5}$ & 23 & 0.064 & 0.124 & $6.7\cdot10^{12}$ & \\

  $10^{2}$  & 42 & 0.076 & 0.101 & $2.7\cdot10^{42}$ & \\
 \hline

\end{tabular}
    \caption{Mixing heuristics for $s_c$}
    \label{tab:cluster_TVDs}
\end{table}

\newpage
\section{Statistics Tables}

\begin{table}[!htbp]
\centering
\begin{tabular}{||l c c||} 
 \hline
 Epsilon & Cluster Runs & Testimony Runs\\ [0.5ex] 
 \hline
  0 & 0.46301799949206396 & 0.43136311539638866  \\
 
 $10^{-3}$ & 0.4048218476793506 & 0.46174179844907726 \\ 

 $10^{-2}$ & 0.48494718152980465 & 0.47074869027270244  \\

 $10^{-1}$ & 0.5014844328223638 & 0.5055669182468262  \\

 $10^{-0.5}$ &  & 0.5186988174796937  \\

 1 & 0.4719985729382476 & 0.5245971456188732  \\

 $10^{0.5}$ & 0.49232776444785387 &  \\

 $10^{1}$ & 0.5323752045233794 &   \\
 
 $10^{1.5}$ & 0.5017081815086254 &   \\

 $10^{2}$ & 0.4903850576917135 & \\  
 \hline
\end{tabular}
    \caption{Correlations Between $s_t$ and $s_c$ (75 million Data Points)}
    \label{tab:cors}
\end{table}

\begin{table}[!htbp]
\centering
\begin{tabular}{||l c c c c c||} 
 \hline
 Epsilon & Maximum & Minimum & Median & First Quartile & Third Quartile\\ 
 \hline
  0 & 0.02358996 & -0.23184358 & -0.10453941 & -0.12250500 & -0.08704500  \\

 $10^{-3}$ & 0.01379449 & -0.22728022 & -0.10521946 & -0.12267225 & -0.08741295 \\ 

 $10^{-2}$ & 0.01592842 & -0.22171994 & -0.10420544 & -0.12195216 & -0.08667915  \\

 $10^{-1}$ & 0.01081431 & -0.23473211 & -0.10578731 & -0.12285399 & -0.08874566  \\

 1 & 0.01980857 & -0.23080875 & -0.10731018 & -0.12490601 & -0.08948498  \\

 $10^{0.5}$ & 0.00619432 & -0.22981396 & -0.11000855 & -0.12752271 & -0.09246568 \\

 $10^{1}$ & -0.00641618 & -0.22835877 & -0.11673956 & -0.13403142 & -0.09957123   \\

 $10^{1.5}$ & -0.01104565 & -0.24685773 & -0.12934086 & -0.14440872 & -0.11445446  \\
 $10^{2}$ & -0.04861434 & -0.22481603 & -0.13583377 & -0.14808502 & -0.12364232\\  
 \hline

\end{tabular}
    \caption{Statistics on $s_t-s_c$ for Cluster Runs}
    \label{tab:s-t for c}
\end{table}

\begin{table}[!htbp]
    \centering
    \begin{tabular}{||l c c c c c||}
 \hline
 Epsilon & Maximum & Minimum & Median & First Quartile & Third Quartile\\ 
 \hline
  0 & 0.02592983 & -0.23254010 & -0.10039906 & -0.11779911 & -0.08318669  \\

 $10^{-3}$ & 0.03517329 & -0.21679741 & -0.10085879 & -0.11799962 & -0.08371213 \\ 

 $10^{-2}$ & 0.01906224 & -0.23425451 & -0.09779002 & -0.11561631 & -0.08032877  \\

 $10^{-1}$ & 0.03492690 & -0.20186416 & -0.08364887 & -0.10066239 & -0.06649066  \\

 $10^{-0.5}$ & 0.04169362 & -0.14930017 & -0.04477747 & -0.06220689 & -0.02841508 \\

 1 & 0.07093255 & -0.11729972 & -0.00425408 & -0.01561957 & 0.00747522  \\
 \hline
    \end{tabular}
    \caption{Statistics on $s_t-s_c$ for Testimony Runs}
    \label{tab:s-c for t}
\end{table}

\begin{table}[!htbp]
    \centering
    \begin{tabular}{||l c c c c c||}
 \hline
 Epsilon & Area & Population & Rel. Internal Nodes & Rel. Internal Edges & Log(Spanning Trees)\\ 
 \hline
  0 & 0.31135174 & 0.52279327 & 0.42121740 & 0.43757661 & 0.51949446  \\

 $10^{-3}$ & 0.31560367 & 0.52251703 & 0.42449003 & 0.44108451 & 0.52192864 \\ 

 $10^{-2}$ & 0.32050756 & 0.52406473 & 0.42808724 & 0.44507853 & 0.52643447 \\

 $10^{-1}$ & 0.31587487 & 0.53135525 & 0.42942184 & 0.44378094 & 0.53844865  \\

 $10^{-0.5}$ & 0.30479716 & 0.41075919 & 0.30929952 & 0.38404223 & 0.49677915 \\

 1 & 0.26316598 & 0.33394644 & 0.21717224 & 0.32861042 & 0.46788565  \\
 \hline
    \end{tabular}
    \caption{Correlations Between Split Frequencies and Various COI Properties for Testimony Runs}
    \label{tab:cor_prop}
\end{table}

\section{Supplementary Correlation Figures}

\begin{figure}[H]
    \centering
    \includegraphics[width=0.725\linewidth]{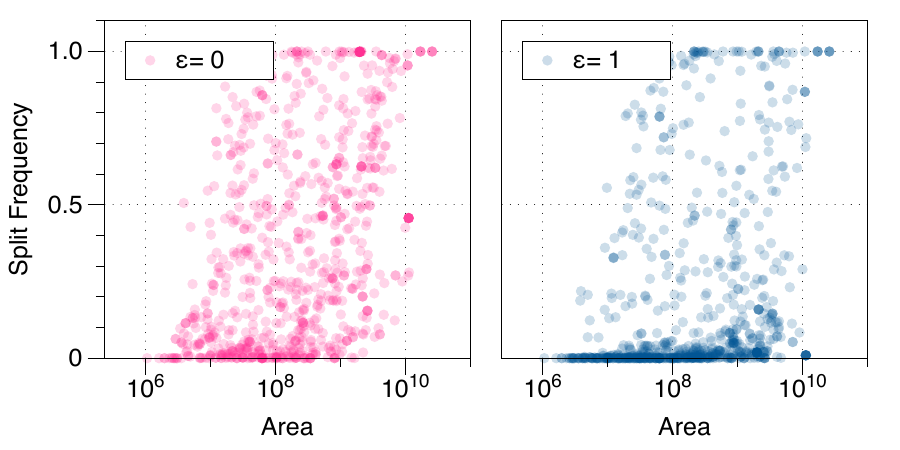}
    \caption{\textbf{Effect of Area on Testimony Splitting: } Scattered testimonies' split frequencies as a function of their area at two privacy budgets. There is a positive correlation between the two values across privacy budgets, indicating that larger testimonies are more likely to be split.}
    \label{fig:area_split}
\end{figure}

\begin{figure}[!htbp]
    \centering
    \includegraphics[width=0.9\linewidth]{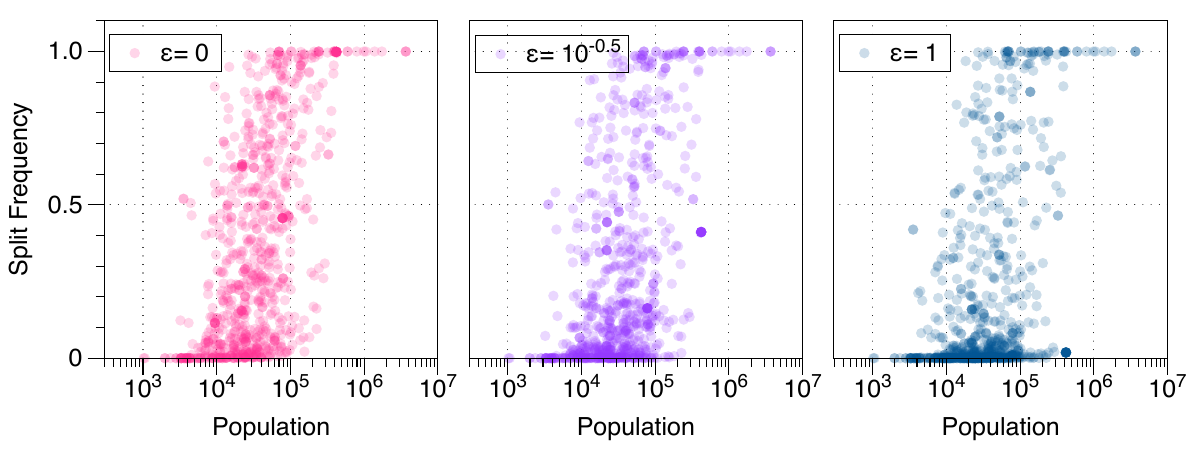}
    \caption{\textbf{Effect of Population on Testimony Splitting: } Scattered testimonies' split frequencies as a function of their population at three privacy budgets. There is a positive correlation between the two values across privacy budgets, indicating that more populous testimonies are more likely to be split. However, this relationship weakens as $\epsilon$ increases.}
    \label{fig:pop_split}
\end{figure}

\end{appendices}
\end{document}